\documentclass[graybox]{svmult}

\usepackage{mathptmx}       
\usepackage{helvet}         
\usepackage{courier}        
\usepackage{type1cm}        
\usepackage{amsmath,amssymb,bm}
\usepackage{makeidx}         
\usepackage{graphicx}        
\usepackage{multicol}        
\usepackage[bottom]{footmisc}

\makeindex             

\def\vp{v_{\rm pole}}
\def\de{\partial}
\def\lm{{\ell m}}

\def\ii{{\rm i}}
\def\l{{\ell }}

\def\I{{\cal I}}
\def\J{{\cal J}}

\def\k{\hat{\hat{k}}}

\begin{document}

\title*{The Effective One Body description of the Two-Body problem}
\author{Thibault Damour and Alessandro Nagar}
\institute{Thibault Damour \at Institut des Hautes Etudes Scientifiques, 35
  route de Chartres, \\ F-91440 Bures-sur-Yvette, France, \email{damour@ihes.fr}
\and Alessandro Nagar \at Institut des Hautes Etudes Scientifiques, 35
  route de Chartres, \\ F-91440 Bures-sur-Yvette, France, 
  \at INFN, Sezione di Torino, Italy, \email{nagar@ihes.fr.}}

\date{today}

\maketitle

\abstract{The Effective One Body (EOB) formalism is an analytical approach 
which aims at providing an accurate description of the motion and radiation 
of coalescing binary black holes with arbitrary mass ratio. 
We review the basic elements of this formalism and discuss its aptitude at 
providing accurate template waveforms to be used for gravitational wave 
data analysis purposes.}

\section{Introduction}
\label{sec:1}

A network of ground-based interferometric gravitational wave (GW) detectors 
(LIGO/VIRGO/GEO/$\ldots$) is currently taking data near its planned
sensitivity~\cite{Sathyaprakash:2009xs}. 
Coalescing black hole binaries are among the most promising, 
and most exciting, GW sources for these detectors. In order to successfully 
detect GWs from coalescing black hole binaries, and to be able to reliably 
measure the physical parameters of the source (masses, spins, $\ldots$), it 
is necessary to know in advance the shape of the GW signals emitted by 
inspiralling and merging black holes. Indeed, the detection and subsequent 
data analysis of GW signals is made by using a large bank of {\it templates} 
that accurately represent the GW waveforms emitted by the source.

\smallskip

Here, we shall introduce the reader to one promising strategy toward having 
an accurate analytical\footnote{Here we use the adjective ``analytical'' for 
methods that solve explicit (analytically given) ordinary differential 
equations (ODE), even if one uses standard (Runge-Kutta-type) numerical 
tools to solve them. The important point is that, contrary to 3D numerical 
relativity simulations, numerically solving ODE's is extremely fast, and can 
therefore be done (possibly even in real time) for a dense sample of 
theoretical parameters, such as orbital ($\nu = m_1 \, m_2 / M , \ldots$) 
or spin ($\hat a_1 = S_1 / Gm_1^2 , \theta_1 , \varphi_1 , \ldots$)
parameters.} description of the motion and radiation of binary black holes, 
which covers all its stages (inspiral, plunge, merger and ring-down): 
the {\it Effective One Body} 
approach~\cite{Buonanno:1998gg,Buonanno:2000ef,Damour:2000we,Damour:2001tu}. 
As early as 2000~\cite{Buonanno:2000ef} this method made several quantitative 
and qualitative predictions concerning the dynamics of the coalescence, and 
the corresponding GW radiation, notably: (i) a blurred transition from
inspiral to a `plunge' that is just a smooth continuation of the inspiral, 
(ii) a sharp transition, around the merger of the black holes, between a 
continued inspiral and a ring-down signal, and (iii) estimates of the radiated 
energy and of the spin of the final black hole. In addition, the effects of 
the individual spins of the black holes were investigated within the 
EOB~\cite{Damour:2001tu,Buonanno:2005xu} and were shown to lead to a larger 
energy release for spins parallel to the orbital angular momentum, and to a 
dimensionless rotation parameter $J/E^2$ always smaller than unity at the 
end of the inspiral (so that a Kerr black hole can form right after the 
inspiral phase). All those predictions have been broadly confirmed by the 
results of the recent numerical simulations performed by several independent 
groups~\cite{Pretorius:2005gq,Pretorius:2006tp,Sperhake:2008ga,Campanelli:2005dd,
Campanelli:2006uy,Campanelli:2006gf,Campanelli:2007cga,Baker:2005vv,
Baker:2006yw, Baker:2006vn,Baker:2007fb,Gonzalez:2006md,
Gonzalez:2007hi,Husa:2007rh,Koppitz:2007ev,Pollney:2007ss, 
Rezzolla:2007xa,Rezzolla:2007rd,Rezzolla:2007rz,Boyle:2006ne,
Boyle:2007ft,Boyle:2008ge,Scheel:2008rj} 
(for a review of numerical relativity results see
also~\cite{Pretorius:2007nq}). 
Note that, in spite of the high computer power used in these simulations, the calculation 
of one sufficiently long waveform (corresponding to specific values of the 
many continuous parameters describing the two arbitrary masses, the initial 
spin vectors, and other initial data) takes on the order of two weeks. This 
is a very strong argument for developing analytical models of waveforms.

\smallskip

Those recent breakthroughs in numerical relativity (NR) open the possibility 
of comparing in detail the EOB description to NR results. This EOB/NR
comparison has been initiated in several 
works~\cite{Buonanno:2006ui,Pan:2007nw,Buonanno:2007pf,
Damour:2007cb,Nagar:2006xv,Damour:2007xr,Damour:2007yf,Damour:2007vq,
Damour:2008te,Damour:2009kr,Buonanno:2009qa}.
The level of analytical/numerical agreement is unprecedented, compared to what
has been previously achieved when comparing other types of analytical
waveforms to numerical ones. In particular,
Refs.~\cite{Damour:2009kr,Buonanno:2009qa} have compared
two different kind of analytical waveforms, 
computed within the EOB framework, to the most accurate
gravitational waveform currently available from the Caltech-Cornell
group, finding that the phase and amplitude differences are
of the order of the numerical error.

If the reader wishes to put the EOB results in contrast with other 
(Post-Newtonian or hybrid) approaches he can consult, 
{\it e.g.}, \cite{Boyle:2007ft,Boyle:2008ge,Baker:2006ha,
Hannam:2007ik,Ajith:2007qp,Ajith:2007kx,Gopakumar:2007vh,Hannam:2007wf}.

\smallskip

Before reviewing some of the technical aspects of the EOB method, let us
indicate some of the historical roots of this method. First, we note that 
the EOB approach comprises three, rather separate, ingredients:
\begin{enumerate}
\item{ a description of the conservative (Hamiltonian) part of the dynamics of two black holes;}
\item{ an expression for the radiation-reaction part of the dynamics;}
\item{ a description of the GW waveform emitted by a coalescing binary system.}
\end{enumerate}

\medskip

For each one of these ingredients, the essential inputs that are used in EOB 
works are high-order post-Newtonian (PN) expanded results which have 
been obtained by many years of work, by many researchers (see references
below). However, one of the key ideas in the EOB philosophy is to avoid using
PN results in their original ``Taylor-expanded'' form ({\it i.e.} $c_0 + c_1 \,
v + c_2 \, v^2 + c_3 \, v^3 + \cdots + c_n \, v^n)$, but to use them instead 
in some {\it resummed} form ({\it i.e.} some non-polynomial function of 
$v$, defined so as to incorporate some of the expected non-perturbative 
features of the exact result). The basic ideas and techniques for resumming 
each ingredient of the EOB are different and have different historical roots. 
Concerning the first ingredient, {\it i.e.} the EOB Hamiltonian, it was inspired 
by an approach to electromagnetically interacting quantum two-body systems 
introduced by Br\'ezin, Itzykson and Zinn-Justin~\cite{Brezin:1970zr}.
 
The resummation of the second ingredient, {\it i.e.} the EOB radiation-reaction 
force ${\mathcal F}$, was originally inspired by the Pad\'e resummation of 
the flux function introduced by Damour, Iyer and Sathyaprakash~\cite{Damour:1997ub}.
Recently, a new and more sophisticated resummation technique for the
radiation reaction force ${\mathcal F}$ has been introduced by 
Damour, Iyer and Nagar~\cite{Damour:2008gu} and further employed in EOB/NR 
comparisons~\cite{Damour:2009kr}. It will be discussed in detail below.

As for the third ingredient, {\it i.e.} the EOB description of the waveform 
emitted by a coalescing black hole binary, it was mainly inspired by the 
work of Davis, Ruffini and Tiomno~\cite{Davis:1972ud} which discovered 
the transition between the plunge signal and a ringing tail when a particle 
falls into a black hole. Additional motivation for the EOB treatment of 
the transition from plunge to ring-down came from work on the, 
so-called, ``close limit approximation''~\cite{Price:1994pm}.

Let us finally note that the EOB approach has been recently 
improved~\cite{Damour:2007yf,Damour:2008gu,Damour:2009kr} by 
following a methodology consisting of studying, element by element, the 
physics behind each feature of the waveform, and on systematically comparing 
various EOB-based waveforms with `exact' waveforms obtained by NR approaches. 
Among these `exact' NR waveforms, it has been useful to consider the 
small-mass-ratio limit~\footnote{Beware that the fonts used in this chapter
make the greek letter $\nu$ (indicating the symmetric mass ratio) look 
very similar to the latin letter $v\neq\nu$ indicating the velocity.} 
$\nu \equiv m_1 \, m_2 / (m_1 + m_2)^2 \ll 1$, in which 
one can use the well controllable `laboratory' of numerical simulations of 
test particles (with an added radiation-reaction force) moving in 
black hole backgrounds \cite{Nagar:2006xv,Damour:2007xr}.

\section{Motion and radiation of binary black holes: 
post-Newtonian expanded results}
\label{sec:2}

Before discussing the various resummation techniques used in the EOB approach, 
let us briefly recall the `Taylor-expanded' results that have been obtained 
by pushing to high accuracies the post-Newtonian (PN) methods.

\smallskip

Concerning the orbital dynamics of compact binaries, we recall that the 
2.5PN-accurate\footnote{As usual `$n$-PN accuracy' means that a result has 
been derived up to (and including) terms which 
are $\sim (v/c)^{2n} \sim (GM/c^2 r)^n$ fractionally smaller than the leading 
contribution.} equations of motion have been derived in the 1980's 
\cite{Damour:1981bh,Damour:1982wm,Schafer:1986rd,kopejkin1985}. Pushing the 
accuracy of the equations of motion to the 3PN ($\sim (v/c)^6$) level proved 
to be a non-trivial task. At first, the representation of black holes by 
delta-function sources and the use of the (non diffeomorphism invariant) 
Hadamard regularization method led to ambiguities in the computation of the 
badly divergent integrals that enter the 3PN equations of 
motion \cite{gr-qc/9712075,gr-qc/0007051}. This problem was solved 
by using the (diffeomorphism invariant) {\it dimensional regularization} 
method ({\it i.e.} analytic continuation in the dimension of space $d$) 
which allowed one to complete the determination of the 3PN-level 
equations of motion \cite{gr-qc/0105038,gr-qc/0311052}. They have also 
been derived by an Einstein-Infeld-Hoffmann-type surface-integral 
approach \cite{IF03}. The 3.5PN terms in the equations of motion 
are also known \cite{gr-qc/0201001,KonigsdorfferFayeSchafer03,gr-qc/0412018}.

\smallskip

Concerning the emission of gravitational radiation, 
two different {\it gravitational-wave generation formalisms} 
have been developed up to a high PN accuracy: 
(i) the Blanchet-Damour-Iyer formalism
\cite{Blanchet:1985sp,Blanchet:1989ki,Damour:1990gj,Damour:1990ji,
Blanchet:1992br,gr-qc/9501030,gr-qc/9710038} 
combines a multipolar post-Minkowskian (MPM) expansion in the 
exterior zone with a post-Newtonian expansion in the near zone; 
while (ii) the Will-Wiseman-Pati formalism
\cite{gr-qc/9608012,gr-qc/9910057,gr-qc/0007087,gr-qc/0201001} 
uses a direct integration of the relaxed Einstein equations. 
These formalisms were used to compute increasingly accurate estimates 
of the gravitational waveforms emitted by inspiralling binaries. 
These estimates include both normal, near-zone generated post-Newtonian 
effects (at the 1PN \cite{Blanchet:1989ki}, 2PN
\cite{gr-qc/9501027,gr-qc/9501029,gr-qc/9608012}, 
and 3PN \cite{gr-qc/0105098,gr-qc/0409094} levels), and more subtle, 
wave-zone generated (linear and non-linear) 
`tail effects' \cite{Blanchet:1992br,W93,BS93,gr-qc/9710038}. However, 
technical problems arose at the 3PN level. 
Similarly to what happened with the equation of motion, 
the representation 
of black holes by `delta-function' sources causes the appearance 
of dangerously divergent integrals in the 3PN multipole moments. 
The use of Hadamard (partie finie) regularization did not allow 
one to unambiguously compute the needed 3PN-accurate quadrupole moment. 
Only the use of the (formally) diffeomorphism-invariant 
{\it dimensional regularization} method allowed one to 
complete the 3PN-level gravitational-radiation 
formalism~\cite{gr-qc/0503044}.

\smallskip

The works mentioned in this Section (see \cite{Blanchet:2002av} 
for a detailed account and more references) finally lead to 
PN-expanded results for the motion and radiation of binary 
black holes. For instance, the 3.5PN equations of motion are 
given in the form ($a=1,2$; $i=1,2,3$)
\begin{equation}
\label{eq6.1}
\frac{d^2 z_a^i}{dt^2} = A_a^{i \, {\rm cons}} + A_a^{iRR} \, ,
\end{equation}
where 
\begin{equation}
\label{eq6.2}
A^{\rm cons} = A_0 + c^{-2} A_2 + c^{-4} A_4 + c^{-6} A_6 \, ,
\end{equation}
denotes the `conservative' 3PN-accurate terms, while
\begin{equation}
\label{eq6.3}
A^{RR} = c^{-5} A_5 + c^{-7} A_7 \, ,
\end{equation}
denotes the time-asymmetric contibutions, linked to `radiation reaction'.

\smallskip

On the other hand, if we consider for simplicity the inspiralling motion 
of a quasi-circular binary system, the essential quantity describing 
the emitted gravitational waveform is the {\it phase} $\phi$ of the 
quadrupolar gravitational wave amplitude 
$h(t) \simeq a(t) \cos (\phi (t) + \delta)$. PN theory allows one to 
derive several different functional expressions for the gravitational 
wave phase $\phi$, as a function either of time or of the instantaneous 
frequency. For instance, as a function of time, $\phi$ admits the 
following explicit expansion in powers of 
$\theta \equiv \nu c^3 (t_c - t) / 5GM$ (where $t_c$ denotes a 
formal `time of coalescence', $M \equiv m_1 + m_2$ and $\nu \equiv m_1 \, m_2 / M^2$)
\begin{equation}
\label{eq6.4}
\phi (t) = \phi_c - \nu^{-1} \, \theta^{5/8} \left( 1 + \sum_{n=2}^7 (a_n + a'_n \, \ln \, \theta) \, \theta^{-n/8} \right) , 
\end{equation}
with some numerical coefficients $a_n , a'_n$ which depend only on 
the dimensionless (symmetric) mass ratio $\nu \equiv m_1 \, m_2 / M^2$. 
The derivation of the 3.5PN-accurate expansion (\ref{eq6.4}) uses both 
the 3PN-accurate conservative acceleration (\ref{eq6.2}) and a 3.5PN
 extension of the (fractionally) 1PN-accurate radiation reaction 
acceleration (\ref{eq6.3}) obtained by assuming a balance between 
the energy of the binary system and the gravitational-wave energy 
flux at infinity (see, {\it e.g.}, \cite{Blanchet:2002av}).

Among the many other possible ways~\cite{Damour:2000zb} of using 
PN-expanded results to predict the GW phase $\phi(t)$, let us mention the semi-analytic 
T4 approximant~\cite{Baker:2006ha,Pan:2007nw}. The GW phase defined 
by the T4 approximant happens to agree well during the inspiral with 
the NR phase in the equal mass case~\cite{Boyle:2007ft}. 
However, this agreement seems to be coincidental
because the T4 phase exhibits significant disagreement with NR results
for other mass ratios~\cite{Damour:2008te} 
(as well as for spinning black-holes~\cite{Hannam:2007wf}).

\section{Conservative dynamics of binary black holes: the Effective One Body approach}\label{sec:3}

The PN-expanded results briefly reviewed in the previous Section are 
expected to yield accurate descriptions of the motion and radiation 
of binary black holes only during their {\it early inspiralling} 
stage, {\it i.e.} as long as the PN expansion parameter $\gamma_e = GM/c^2 R$ 
(where $R$ is the distance between the two black holes) stays significantly 
smaller than the value $\sim \frac{1}{6}$ where the orbital motion is 
expected to become dynamically unstable (`last stable circular orbit' 
and beginning of a `plunge' leading to the merger of the two black holes). 
One needs a better description of the motion and radiation to describe 
the {\it late inspiral} (say $\gamma_e \gtrsim \frac{1}{12}$), as well 
as the subsequent {\it plunge} and {\it merger}. One possible strategy 
for having a complete description of the motion and radiation of 
binary black holes, covering all the stages (inspiral, plunge, 
merger, ring-down), would then be to try to `stitch together' 
PN-expanded analytical results describing the early inspiral
 phase with 3D numerical results describing the end of the 
inspiral, the plunge, the merger and the ring-down of the 
final black hole, see, {\it e.g.}, Refs.~\cite{Baker:2006kr,Pan:2007nw}.

\smallskip

However, we wish to argue that the EOB approach makes a better use of all 
the analytical information contained in the PN-expanded results 
(\ref{eq6.1})-(\ref{eq6.3}). The basic claim (first made 
in~\cite{Buonanno:1998gg,Buonanno:2000ef}) is that the use of suitable 
{\it resummation methods} should allow one to describe, by analytical 
tools, a {\it sufficiently accurate} approximation of the {\it entire waveform}, 
from inspiral to ring-down, including the non-perturbative plunge 
and merger phases. To reach such a goal, one needs to make use of 
several tools: (i) resummation methods, (ii) exploitation of the flexibility 
of analytical approaches, (iii) extraction of the non-perturbative information 
contained in various numerical simulations, (iv) qualitative understanding 
of the basic physical features which determine the waveform.

\smallskip

Let us start by discussing the first tool used in the EOB approach: the 
systematic use of resummation methods. 
Essentially two resummation methods have been employed (and combined) and some evidence 
has been given that they do significantly improve the convergence 
properties of PN expansions. 
The first method is the 
systematic use of {\it Pad\'e approximants}. 
It has been shown in Ref.~\cite{Damour:1997ub} that near-diagonal Pad\'e 
approximants of the radiation reaction force\footnote{We henceforth denote 
by ${\mathcal F}$ the {\it Hamiltonian} version of the radiation reaction 
term $A^{RR}$, Eq.~(\ref{eq6.3}), in the (PN-expanded) equations of motion. 
It can be heuristically computed up to (absolute) 5.5PN 
\cite{gr-qc/0105098, gr-qc/0406012,gr-qc/0503044} and even 
6PN~\cite{gr-qc/0105099} 
order by assuming that the energy radiated in gravitational waves 
at infinity is balanced by a loss of the dynamical energy of the binary
system.} ${\mathcal F}$ seemed to provide a good representation of 
${\mathcal  F}$ down to the last stable orbit (which is expected to occur 
when $R \sim 6GM/c^2$, {\it i.e.} when $\gamma_e \simeq \frac{1}{6}$). 
In addition, a new route to the resummation of ${\mathcal F}$
has been proposed very recently in Ref.~\cite{Damour:2008gu}. 
This approach, that will be discussed in detail below, is based 
on a  new multiplicative decomposition of the metric multipolar
waveform (which is originally given as a standard PN series). 
In this case, Pad\'e approximants prove to be useful 
to further improve the convergence properties of one particular
factor of this multiplicative decomposition.

The second resummation method is a novel approach to the dynamics 
of compact binaries, which constitutes the core of the Effective 
One Body (EOB) method. 

\smallskip

For simplicity of exposition, let us first explain the EOB method at the 
2PN level. The starting point of the method is the 2PN-accurate Hamiltonian 
describing (in Arnowitt-Deser-Misner-type coordinates) the conservative, or 
time symmetric, part of the equations of motion (\ref{eq6.1}) ({\it i.e.} 
the truncation $A^{\rm cons} = A_0 + c^{-2} A_2 + c^{-4} A_4$ of 
Eq.~(\ref{eq6.2})) say $H_{\rm 2PN} ({\bm q}_1 - {\bm q}_2 , {\bm p}_1 , {\bm
  p}_2)$. 
By going to the center of mass of the system $({\bm p}_1 + {\bm p}_2 = 0)$, 
one obtains a PN-expanded Hamiltonian describing the {\it relative motion}, 
${\bm q} = {\bm q}_1 - {\bm q}_2$, ${\bm p} = {\bm p}_1 = - {\bm p}_2$:
\begin{equation}
\label{eq7.1}
H_{\rm 2PN}^{\rm relative} ({\bm q} , {\bm p}) = H_0 ({\bm q} , 
{\bm p}) + \frac{1}{c^2} \, H_2 ({\bm q} , {\bm p}) + \frac{1}{c^4} \, H_4 ({\bm q} , {\bm p}) \, ,
\end{equation}
where $H_0 ({\bm q} , {\bm p}) = \frac{1}{2\mu} \, {\bm p}^2 + \frac{GM\mu}
{\vert {\bm q} \vert}$ (with $M \equiv m_1 + m_2$ and $\mu = m_1 \, m_2 / M$) 
corresponds to the Newtonian approximation to the relative motion, while $H_2$ 
describes 1PN corrections and $H_4$ 2PN ones. It is well known that, at the 
Newtonian approximation, $H_0 ({\bm q} , {\bm p})$ can be thought of as 
describing a `test particle' of mass $\mu$ orbiting around an `external 
mass' $GM$. The EOB approach is a {\it general relativistic generalization} 
of this fact. It consists in looking for an `external spacetime geometry'
 $g_{\mu\nu}^{\rm ext} (x^{\lambda} ; GM)$ such that the geodesic dynamics of 
a `test particle' of mass $\mu$ within $g_{\mu\nu}^{\rm ext} (x^{\lambda} , 
GM)$ is {\it equivalent}  (when expanded in powers of $1/c^2$) to the
original, relative PN-expanded dynamics (\ref{eq7.1}). 

\smallskip 

Let us explain the idea, proposed in \cite{Buonanno:1998gg}, for establishing 
a `dictionary' between the real relative-motion dynamics, (\ref{eq7.1}), and 
the dynamics of an `effective' particle of mass $\mu$ moving in 
$g_{\mu\nu}^{\rm ext} (x^{\lambda} , GM)$. The idea consists in `thinking 
quantum mechanically'\footnote{This is related to an idea emphasized many 
times by John Archibald Wheeler: quantum mechanics can often help us in 
going to the essence of classical mechanics.}. Instead of thinking in terms 
of a classical Hamiltonian, $H({\bm q}, {\bm p})$ 
(such as $H_{\rm 2PN}^{\rm relative}$, Eq.~(\ref{eq7.1})), and of its classical 
bound orbits, we can think in terms of the quantized energy levels $E(n,\ell)$ 
of the quantum bound states of the Hamiltonian operator $H (\hat{\bm q} ,\hat{\bm p})$. 
These energy levels will depend on two (integer valued) quantum numbers 
$n$ and $\ell$. Here (for a spherically symmetric interaction, as appropriate 
to $H^{\rm relative}$), $\ell$ parametrizes the total orbital angular 
momentum (${\bm L}^2 = \ell (\ell + 1) \, \hbar^2$), while $n$ represents 
the `principal quantum number' $n = \ell + n_r + 1$, where $n_r$ 
(the `radial quantum number') denotes the number of nodes in the radial 
wave function. The third `magnetic quantum number' $m$ 
(with $-\ell \leq m \leq \ell$) does not enter the energy levels because 
of the spherical symmetry of the two-body interaction (in the center of of 
mass frame). For instance, a non-relativistic Coulomb (or Newton!) 
interaction
\begin{equation}
\label{eqn1}
H_0 = \frac{1}{2\mu} \, {\bm p}^2 + \frac{GM\mu}{\vert {\bm q} \vert}
\end{equation}
gives rise to the well-known result
\begin{equation}
\label{eqn2}
E_0 (n,\ell) = - \frac{1}{2} \, \mu \left(\frac{GM\mu}{n \, \hbar} \right)^2 \, ,
\end{equation}
which depends only on $n$ (this is the famous Coulomb degeneracy). 
When considering the PN corrections to $H_0$, as in Eq.~(\ref{eq7.1}), 
one gets a more complicated expression of the form
\begin{equation}
\label{eqn3}
E_{\rm 2PN}^{\rm relative} (n,\ell) = - \frac{1}{2} \mu  
\frac{\alpha^2}{n^2} \biggl[ 1 + 
\frac{\alpha^2}{c^2} \left( \frac{c_{11}}{n\ell} + \frac{c_{20}}{n^2} \right)
+ \frac{\alpha^4}{c^4} \left( \frac{c_{13}}{n\ell^3} + \frac{c_{22}}{n^2 \ell^2} + \frac{c_{31}}{n^3 \ell} + \frac{c_{40}}{n^4} \right)\biggl] \, ,
\end{equation}
where we have set $\alpha \equiv GM\mu / \hbar = G \, m_1 \, m_2 / \hbar$, 
and where we consider, for simplicity, the (quasi-classical) limit 
where $n$ and $\ell$ are large numbers. The 2PN-accurate result (\ref{eqn3}) 
had been derived by Damour and Sch\"afer \cite{Damour:1988mr} 
as early as 1988. The dimensionless coefficients $c_{pq}$ are 
functions of the symmetric mass ratio $\nu \equiv \mu / M$, for 
instance $c_{40} = \frac{1}{8} (145 - 15\nu + \nu^2)$. 
In classical mechanics ({\it i.e.} for large $n$ and $\ell$), it 
is called the `Delaunay Hamiltonian', {\it i.e.} the Hamiltonian 
expressed in terms of the 
{\it action variables}\footnote{We consider, for simplicity, 
`equatorial' motions with $m=\ell$, {\it i.e.}, classically, 
$\theta = \frac{\pi}{2}$.} $J = \ell \hbar 
= \frac{1}{2\pi} \oint p_{\varphi} \,d\varphi$, 
and $N = n \hbar = I_r + J$, with $I_r = \frac{1}{2\pi} \oint p_r \, dr$.

\smallskip

The energy levels (\ref{eqn3}) encode, in a gauge-invariant way, 
the 2PN-accurate relative dynamics of a `real' binary. Let us 
now consider an auxiliary problem: the `effective' dynamics of 
one body, of mass $\mu$, following a geodesic in some `external' 
(spherically symmetric) metric\footnote{It is convenient to write 
the `external metric' in Schwarzschild-like coordinates. 
Note that the external radial coordinate $R$ differs from 
the two-body ADM-coordinate relative distance 
$R^{\rm ADM} = \vert {\bm q} \vert$. The transformation 
between the two coordinate systems has been determined 
in Refs.~\cite{Buonanno:1998gg,Damour:2000we}.}
\begin{equation}
\label{eq7.2}
g_{\mu\nu}^{\rm ext} \, dx^{\mu} \, dx^{\nu} = - A(R) \, c^2 \, d T^2 + B(R) \, d R^2 + R^2 (d\theta^2 + \sin^2 \theta \, d \varphi^2) \, .
\end{equation}
Here, the {\it a priori unknown} metric functions $A(R)$ and $B(R)$ 
will be constructed in the form of expansions in $GM/c^2 R$:
\begin{eqnarray}
\label{eqn4}
A(R) &= &1 + a_1 \, \frac{GM}{c^2 R} + a_2 \left( \frac{GM}{c^2 R} \right)^2 + a_3 \left( \frac{GM}{c^2 R} \right)^3 + \cdots \, ; \nonumber \\
B(R) &= &1 + b_1 \, \frac{GM}{c^2 R} + b_2 \left( \frac{GM}{c^2 R} \right)^2 + \cdots \, ,
\end{eqnarray}
where the dimensionless coefficients $a_n , b_n$ depend on $\nu$. 
From the Newtonian limit, it is clear that we should set $a_1 = -2$. 
By solving (by separation of variables) the `effective'
 Hamilton-Jacobi equation
$$
g_{\rm eff}^{\mu\nu} \, \frac{\partial S_{\rm eff}}{\partial x^{\mu}} \, \frac{\partial S_{\rm eff}}{\partial x^{\nu}} + \mu^2 c^2 = 0 \, ,
$$
\begin{equation}
\label{eqn5}
S_{\rm eff} = - {\mathcal E}_{\rm eff} \, t + J_{\rm eff} \, \varphi + S_{\rm eff} (R) \, ,
\end{equation}
one can straightforwardly compute (in the quasi-classical, large quantum
numbers limit) the Delaunay 
Hamiltonian ${\mathcal E}_{\rm eff} (N_{\rm eff} , J_{\rm eff})$, 
with $N_{\rm eff} = n_{\rm eff} \, \hbar$, $J_{\rm eff} = \ell_{\rm eff} \,
\hbar$ (where $N_{\rm eff} = J_{\rm eff} + I_R^{\rm eff}$, with $I_R^{\rm eff}
= \frac{1}{2\pi} \oint p_R^{\rm eff} 
\, dR$, $P_R^{\rm eff} = \partial S_{\rm eff} (R) / dR$). 
This  yields a result of the form
\begin{align}
{\mathcal E}_{\rm eff} (n_{\rm eff},\ell_{\rm eff}) = \mu c^2 - \frac{1}{2} \,
\mu \ \frac{\alpha^2}{n_{\rm eff}^2}
\biggl[ 1 &+ \frac{\alpha^2}{c^2} \left( \frac{c_{11}^{\rm eff}}{n_{\rm eff}
    \ell_{\rm eff}} + \frac{c_{20}^{\rm eff}}{n_{\rm eff}^2} \right)\nonumber\\
&+ \frac{\alpha^4}{c^4} \left( \frac{c_{13}^{\rm eff}}{n_{\rm eff} \ell_{\rm eff}^3} 
+ \frac{c_{22}^{\rm eff}}{n_{\rm eff}^2 \ell_{\rm eff}^2} + \frac{c_{31}^{\rm eff}}{n_{\rm eff}^3 \ell_{\rm eff}} 
+ \frac{c_{40}^{\rm eff}}{n_{\rm eff}^4} \right)\biggl] ,
\label{eqn6}
\end{align}
where the dimensionless coefficients $c_{pq}^{\rm eff}$ are now 
functions of the unknown coefficients $a_n , b_n$ entering the 
looked for `external' metric coefficients (\ref{eqn4}).

\smallskip

\begin{figure}[t]
\begin{center}
\includegraphics[height=8cm ]{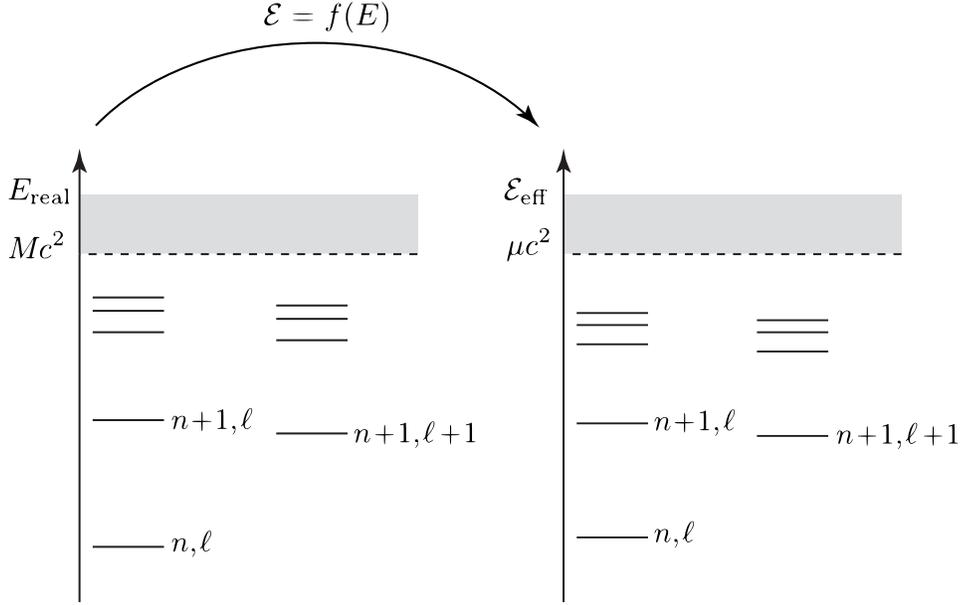}
\caption{\label{fig:1}Sketch of the correspondence between the quantized energy levels of the 
real and effective conservative dynamics. $n$ denotes the `principal 
quantum number' ($n = n_r + \ell + 1$, with $n_r = 0,1,\ldots$ denoting the
number 
of nodes in the radial function), while $\ell$ denotes the (relative) orbital 
angular momentum $({\bm L}^2 = \ell (\ell + 1) \, \hbar^2)$. Though the EOB 
method is purely classical, it is conceptually useful to think in terms of 
the underlying (Bohr-Sommerfeld) quantization conditions of the action 
variables $I_R$ and $J$ to motivate the identification between $n$ and 
$\ell$ in the two dynamics.}
\end{center}
\end{figure}

At this stage, one needs (as in the famous AdS/CFT correspondence) to define a
`dictionary' 
between the real (relative) two-body dynamics, summarized in Eq.~(\ref{eqn3}),
and 
the effective one-body one, summarized in Eq.~(\ref{eqn6}). As, on both sides, 
quantum mechanics tells us that the action variables are quantized in integers 
($N_{\rm real} = n \hbar$, $N_{\rm eff} = n_{\rm eff} \hbar$, etc.) it is most 
natural to identify $n=n_{\rm eff}$ and $\ell = \ell_{\rm eff}$. One then
still 
needs a rule for relating the two different energies $E_{\rm real}^{\rm
  relative}$ 
and ${\mathcal E}_{\rm eff}$. Ref.~\cite{Buonanno:1998gg} proposed to look for
a 
general map between the real energy levels and the effective ones (which, as
seen 
when comparing (\ref{eqn3}) and (\ref{eqn6}), cannot be directly identified 
because they do not include the same rest-mass 
contribution\footnote{Indeed $E_{\rm real}^{\rm total} = Mc^2 + E_{\rm
    real}^{\rm relative} 
= Mc^2 + \mbox{Newtonian terms} + {\rm 1PN} / c^2 + \cdots$, 
while ${\mathcal E}_{\rm effective} = \mu c^2 + N + {\rm 1PN} / c^2 +
\cdots$.}), namely
\begin{equation}
\label{eqn7}
\frac{{\mathcal E}_{\rm eff}}{\mu c^2} - 1 = f \left( \frac{E_{\rm real}^{\rm relative}}{\mu c^2} \right) = \frac{E_{\rm real}^{\rm relative}}{\mu c^2} \left( 1 + \alpha_1 \, \frac{E_{\rm real}^{\rm relative}}{\mu c^2} + \alpha_2 \left( \frac{E_{\rm real}^{\rm relative}}{\mu c^2} \right)^2 + \cdots \right) \, .
\end{equation}
The `correspondence' between the real and effective energy levels is illustrated in Fig.~\ref{fig:1}

Finally, identifying ${\mathcal E}_{\rm eff} (n,\ell) / \mu c^2$ to $f (E_{\rm real}^{\rm relative} / \mu c^2)$ 
yields six equations, relating the six coefficients $c_{pq}^{\rm eff} (a_2 ,
  a_3 ; b_1 , b_2)$ 
to the six $c_{pq} (\nu)$ and to the two energy coefficients $\alpha_1$ and
$\alpha_2$. It is natural to set $b_1 = + 2$ (so that the linearized effective
metric coincides with the linearized Schwarzschild metric with 
mass $M = m_1 + m_2$). One then finds that there exists a {\it unique} solution 
for the remaining five unknown coefficients $a_2 , a_3 , b_2 , \alpha_1$ and $\alpha_2$. 
This solution is very simple:
\begin{equation}
\label{eqn8}
a_2 = 0  \, , \quad a_3 = 2 \nu \, , \quad b_2 = 4 - 6 \nu \, , \quad \alpha_1 = \frac{\nu}{2} \, , \quad \alpha_2 = 0 \, .
\end{equation}
Note, in particular, that the map between the two energies is simply
\begin{equation}
\label{eqn9}
\frac{{\mathcal E}_{\rm eff}}{\mu c^2} = 1 + \frac{E_{\rm real}^{\rm
    relative}}{\mu c^2} 
\left( 1 + \frac{\nu}{2} \, \frac{E_{\rm real}^{\rm relative}}{\mu c^2}\right) 
= \frac{s - m_1^2 \, c^4 - m_2^2 \, c^4}{2 \, m_1 \, m_2 \, c^4}
\end{equation}
where $s = ({\mathcal E}_{\rm real}^{\rm tot})^2 \equiv (M c^2 + E_{\rm  real}^{\rm relative})^2$ 
is Mandelstam's invariant $=-(p_1+p_2)^2$. Note also that, at 2PN accuracy, the crucial 
`$g_{00}^{\rm ext}$' metric coefficient $A(R)$ (which fully encodes the 
energetics of circular orbits) is given by the remarkably simple PN 
expansion
\begin{equation}
\label{eq7.3}
A_{\rm 2PN} (R) = 1-2u + 2 \, \nu \, u^3 \, ,
\end{equation}
where $u \equiv GM/(c^2 R)$ and $\nu \equiv \mu / M \equiv m_1 \, m_2 / (m_1 + m_2)^2$.

The dimensionless parameter $\nu \equiv \mu / M$ varies between $0$ (in the
test mass 
limit $m_1 \ll m_2$) and $\frac{1}{4}$ (in the equal-mass case $m_1 = m_2$). 
When $\nu \to 0$, Eq.~(\ref{eq7.3}) yields back, as expected, the well-known 
Schwarzschild time-time metric coefficient 
$- g_{00}^{\rm Schw} = 1 - 2u = 1 - 2GM / c^2 R$. One therefore sees 
in Eq.~(\ref{eq7.3}) the r\^ole of $\nu$ as a {\it deformation parameter} 
connecting a well-known test-mass result to a non trivial and new 
2PN result. It is also to be noted that the 1PN EOB result $A_{\rm 1PN} (R) =
1-2u$ 
happens to be $\nu$-independent, and therefore identical to $A^{\rm Schw} =
1-2u$. This is remarkable in view of the many non-trivial $\nu$-dependent 
terms in the 1PN relative dynamics. The physically real 1PN $\nu$-dependence 
happens to be fully encoded in the function $f(E)$ mapping the two energy 
spectra given in Eq.~(\ref{eqn9}) above.

Let us emphasize the remarkable simplicity of the 2PN result (\ref{eq7.3}). 
The 2PN Hamiltonian (\ref{eq7.1}) contains eleven rather complicated 
$\nu$-dependent terms. After transformation to the EOB format, the dynamical 
information contained in these eleven coefficients gets {\it condensed} 
into the very simple additional contribution $+ \, 2 \, \nu \, u^3$ in $A(R)$, 
together with an equally simple contribution in the radial metric coefficient: 
$(A(R) \, B(R))_{\rm 2PN} = 1 - 6 \, \nu \, u^2$. This condensation process 
is even more drastic when one goes to the next (conservative) post-Newtonian 
order: the 3PN level, i.e. additional terms of order ${\mathcal O} (1/c^6)$ in 
the Hamiltonian (\ref{eq7.1}). As mentioned above, the complete obtention of 
the 3PN dynamics has represented quite a theoretical challenge and the final, 
resulting Hamiltonian is quite complicated. Even after going to the center of 
mass frame, the 3PN additional contribution $\frac{1}{c^6} \, H_6 ({\bm q} ,
{\bm p})$ to Eq.~(\ref{eq7.1}) introduces eleven new complicated $\nu$-dependent
coefficients. 
After transformation to the EOB format \cite{Damour:2000we}, these eleven new 
coefficients get ``condensed'' into only {\it three} additional terms: 
(i) an additional contribution to $A(R)$, (ii) an additional contribution 
to $B(R)$, and (iii) a ${\mathcal O} ({\bm p}^4)$ modification of the `external' 
geodesic Hamiltonian. For instance, the crucial 3PN $g_{00}^{\rm ext}$ metric 
coefficient becomes
\begin{equation}
\label{eq7.5}
A_{\rm 3PN} (R) = 1-2u + 2 \, \nu \, u^3 + a_4 \, \nu \, u^4 \, ,
\end{equation}
where $u=GM/(c^2 R)$, 
\begin{equation}
\label{eq7.6}
a_4 = \frac{94}{3} - \frac{41}{32} \, \pi^2 \simeq 18.6879027 \ ,
\end{equation}
while the  additional contribution to $B(R)$ gives
\begin{equation}
\label{eq:D}
D_{\rm 3PN}(R)\equiv(A(R)B(R))_{\rm 3PN} = 1-6\nu u^2+2(3\nu-26)\nu u^3 \ .
\end{equation}
Remarkably, it is found that the very simple 2PN energy map 
Eq.~(\ref{eqn9}) does not need to be modified at the 3PN level.

\begin{figure}[t]
\centering
\includegraphics[height=7 cm]{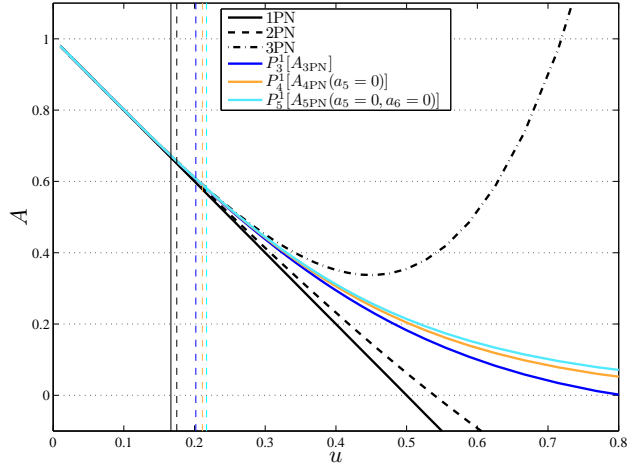}
\caption{Various approximations and Pad\'e resummation of the EOB radial
potential $A(u)$, where $u=GM/(c^2R)$, for the equal-mass case $\nu=1/4$. 
The vertical dashed lines indicate the corresponding 
(adiabatic) LSO location~\cite{Buonanno:1998gg} 
defined by the condition 
$d^2{\cal E}^0_{\rm eff}/dR^2=d{\cal E}_{\rm eff}^0/dR=0$,
where ${\cal E}^0_{\rm eff}$ is the effective energy along
the sequence of circular orbits ({\it i.e.}, when $P_{R}^{\rm eff}=0$). 
\label{fig:fig2}}       
\end{figure}

The fact that the 3PN coefficient $a_4$ in the crucial 
`effective radial potential' $A_{\rm 3PN} (R)$, Eq.~(\ref{eq7.5}), 
is rather large and positive indicates that the $\nu$-dependent 
nonlinear gravitational effects lead, for comparable 
masses $(\nu \sim \frac{1}{4}$), to a last stable 
(circular) orbit (LSO) which has a higher frequency and 
a larger binding energy than what a naive scaling from 
the test-particle limit $(\nu \to 0)$ would suggest. 
Actually, the PN-expanded form (\ref{eq7.5}) of $A_{\rm 3PN} (R)$ 
does not seem to be a good representation of the (unknown) exact 
function $A_{\rm EOB} (R)$ when the (Schwarzschild-like) relative 
coordinate $R$ becomes smaller than about $6 GM / c^2$ (which is 
the radius of the LSO in the test-mass limit). 
In fact, by continuity with the test-mass case, one a priori
expects that $A_{\rm 3PN}(R)$ always exhibits a simple zero 
defining an EOB ``effective horizon'' that is smoothly connected
to the Schwarzschild event horizon at $R= 2GM/c^2$ when $\nu\to 0$. 
However, the large value of the $a_4$ coefficient does actually 
prevent $A_{\rm 3PN}$ to have this property
when $\nu$ is too large, and in particular when $\nu=1/4$, 
as it is visually explained in Fig.~\ref{fig:fig2}.
The black curves in the figure represent the $A$ function 
at 1PN (solid line), 2PN (dashed line) and 3PN (dash-dot line) 
approximation: while the 2PN curve still has a simple zero, 
the 3PN does not, due to the large value of $a_4$. 
It was therefore suggested~\cite{Damour:2000we} 
to further resum\footnote{The PN-expanded EOB 
building blocks $A(R) , B(R) , \ldots$ already 
represent a {\it resummation} of the PN dynamics 
in the sense that they have ``condensed'' the many 
terms of the original PN-expanded Hamiltonian within 
a very concise format. But one should not refrain to 
further resum the EOB building blocks themselves, if 
this is physically motivated.} $A_{\rm 3PN} (R)$ by 
replacing it by a suitable Pad\'e $(P)$ approximant. 
For instance, the replacement of $A_{\rm 3PN} (R)$ by
\begin{equation}
\label{eq7.7}
A_3^1 (R) \equiv P_3^1 [A_{\rm 3PN} (R)] = \frac{1+n_1 u}{1+d_1 u + d_2 u^2 + d_3 u^3}
\end{equation}
ensures that the $\nu = \frac{1}{4}$ case is smoothly 
connected with the $\nu = 0$ limit, as Fig.~\ref{fig:fig2}
clearly shows\footnote{We recall that the coefficient $n_1$ 
and $(d_1,d_2,d_3)$ of the Pad\'e approximant are determined 
by the condition that the first four terms of the Taylor expansion 
of $A_3^1$ in powers of $u=GM/(c^2R)$ coincide with $A_{\rm 3PN}$.}.

The use of Eq.~(\ref{eq7.7}) was suggested before one had any 
(reliable) non-perturbative information on the binding of 
close black hole binaries. Later, a comparison with 
some ``waveless'' numerical simulations of circular black 
hole binaries~\cite{Damour:2002qh} has given some evidence 
that Eq.~(\ref{eq7.7}) is physically adequate. 
In Refs.~\cite{Damour:2001tu,Damour:2002qh} it was also 
emphasized that, in principle, the comparison between 
numerical data and EOB-based predictions should allow 
one to determine the effect of the unknown higher PN 
contributions to Eq.~(\ref{eq7.5}). For instance, one 
can add a 4PN-like term $+ a_5 \nu u^5$ or a 5PN-like term
$+a_6 \nu u^6$ in Eq.~(\ref{eq7.5}), and then Pad\'e the 
resulting radial function. The new {\it resummed} $A$
potential will exhibit an explicit dependence 
on $a_5$ (at 4PN) or $(a_5,a_6)$ (at 5PN), 
that is 
\begin{equation}
\label{A_4PN}
A^1_4(R;a_5,\nu) = P^{1}_{4}\left[A_{\rm 3PN}(R) + \nu a_5 u^5 \right],
\end{equation}
or 
\begin{equation}
\label{A_5PN}
A^1_5(R;a_5,a_6,\nu) = P^{1}_{5}\left[A_{\rm 3PN}(R) + \nu a_5 u^5 + \nu a_6 u^6\right].
\end{equation}
Comparing the predictions of $A_4^1(R;a_5,\nu)$ or $A^1_5(R;a_5,a_6,\nu)$ 
to numerical data might then determine what is the physically 
preferred ``effective'' value of the unknown 
coefficient $a_5$ (if working at 4PN effective accuracy) 
or of the doublet $(a_5,a_6)$ (when including also 5PN corrections).
For illustrative purposes, Fig.~\ref{fig:fig2} shows the effect of the Pad\'e
resummation with $a_5=a_6=0$ and $\nu=1/4$. 
Note that the Pad\'e resummation procedure is injecting some ``information''
beyond that contained in the numerical  values of the PN expansion
coefficients $a_n$'s of $A(R)$. As a consequence, the operation of 
Pad\'eing and of restricting $a_5$ and $a_6$ to the (3PN-compatible)
values $a_5=0=a_6$ do not commute: $A^1_4(R;0,1/4)\neq A^1_5(R;0,0,1/4)\neq A^1_3(R,1/4)$.
In this respect, let us also mention that 
the 4PN $a_5$-dependent 
Pad\'e approximant $A^1_4(R;a_5,\nu)$ 
exactly reduces to the 3PN Pad\'e approximant $A^1_3(R;\nu)$ when
$a_5$ is replaced by the following function of $\nu$
\begin{equation}
a_5^{\rm 3PN}(\nu)\equiv \dfrac{\nu(3392-123\pi^2)^2}{18432(\nu-4)}.
\end{equation}
Note that the value of the $A^1_3$-reproducing effective 4PN coefficient
$a_5^{\rm 3PN}(\nu)$ in the equal mass case is 
$a_5^{\rm 3PN}(1/4)\simeq -17.158031$. This is numerically compatible 
with the value $a_5=-17.16$ quoted in Ref.~\cite{Boyle:2008ge}
(but note that the correct $A^1_3$-reproducing 4PN coefficient depends
on the symmetric mass ratio $\nu$).
Similarly, when working at the 5PN level, 
$A^1_5(R;a_5,a_6,\nu)$ exactly reduces to the 4PN Pad\'e approximant
$A^1_4(R;a_5,\nu)$ when $a_6$ is replaced by the following function
of both $\nu$ and $a_5$:
\begin{align}
 & a_6^{\rm 4PN}(\nu,a_5)\equiv\nonumber\\
&\frac{\nu  \left(2304 a_5^2+96 \left(3392-123 \pi ^2\right)
  a_5+\left(3776-123 \pi ^2\right) \left(32 (3 \nu +94) -123\pi^2 \right)\right)}{24 \left[\left(3776-123 \pi ^2\right) \nu -1536\right]}.
\end{align}
The use of numerical relativity data to 
constrain the values of the higher PN parameters $(a_5,a_6)$ is 
an example of the useful {\it flexibility}~\cite{gr-qc/0211041}  
of analytical approaches: the fact that one can tap 
numerically-based, non-perturbative information to 
improve the EOB approach. 
The flexibility of the EOB approach related to the use of
the $a_5$-dependent radial potential
$A_4^1(R;a_5,\nu)$ has been exploited in several recent 
works~\cite{Buonanno:2007pf,Damour:2007yf,Damour:2007vq,Damour:2008te,Boyle:2008ge,Buonanno:2009qa}
focusing on the comparison of EOB-based waveforms with
waveforms computed via numerical relativity simulations.
Collectively,
all these studies have shown that it is possible to constrain
$a_5$ (together with other flexibility parameters related to
the resummation of radiation reaction, see below) so as to yield 
an excellent agreement (at the level of the published 
numerical errors) between EOB and numerical relativity waveforms.
The result, however, cannot be summarized by stating that $a_5$ 
is constrained to be in the vicinity of a special numerical value.
Rather, one finds a strong correlation between $a_5$ and other
parameters, notably the radiation reaction parameter $v_{\rm pole}$
introduced below.
More recently, Ref.~\cite{Damour:2009kr} could get rid of the flexibility
parameters (such as $v_{\rm pole}$) related to the resummation
of radiation reaction, and has shown that one can get an excellent
agreement with numerical relativity data by
using {\it only} the flexibility in the 
doublet $(a_5,a_6)$ (the other parameters being essentially fixed
internally to the formalism). We shall discuss this result
further in Sec.~\ref{sec:5} below.

The same kind of $\nu$-continuity argument discussed so far 
for the $A$ function needs to be applied also to the 
$D(R)_{\rm 3PN}$ function defined in Eq.~\eqref{eq:D}. 
A straightforward way to ensure that the $D$ function 
stays positive when $R$ decreases (since it is $D=1$ when $\nu\to 0$) 
is to replace $D_{\rm 3PN}(R)$ by 
$D^0_3(R)\equiv P^0_3\left[D_{\rm 3PN}(R)\right]$,
where $P^0_3$ indicates the $(0,3)$ Pad\'e approximant and
explicitly reads
\begin{equation}
D^0_3(R)=\dfrac{1}{1+6\nu u^2  -2(3\nu-26)\nu u^3}.
\end{equation}
The resummation of $A$ (via Pad\'e approximants) is necessary 
for ensuring the existence and $\nu$-continuity of a 
{\it last stable orbit} (see vertical lines in Fig.~\ref{fig:fig2}), 
as well as the existence and $\nu$-continuity 
of a {\it last unstable orbit}, i.e. of a $\nu$-deformed
analog of the light ring $R=3GM/c^2$ when $\nu\to 0$. 
We recall that, when $\nu=0$, the light ring corresponds to the 
circular orbit of a massless particle, 
or of an extremely relativistic massive particle, and is technically 
defined by looking for the maximum of $A(R)/R^2$, i.e. by 
solving $ (d/dR) (A(R)/R^2) =0$. When $\nu\neq 0$ and when considering
the quasi-circular plunge following the crossing of the last stable
orbit, the ``effective'' meaning of the ``$\nu$-deformed light ring''
(technically defined by solving $ (d/dR) (A(R:\nu)/R^2) =0$) is to
entail, in its vicinity, the existence of a maximum of the orbital
frequency $\Omega=d\varphi/dt$ (the resummation of $D(R)$ plays
a useful role in ensuring the $\nu$-continuity of this plunge 
behavior). 

\section{Description of radiation-reaction effects in 
the Effective One Body approach}
\label{sec:4}

\begin{figure}[t]
\centering
\includegraphics[height=7 cm]{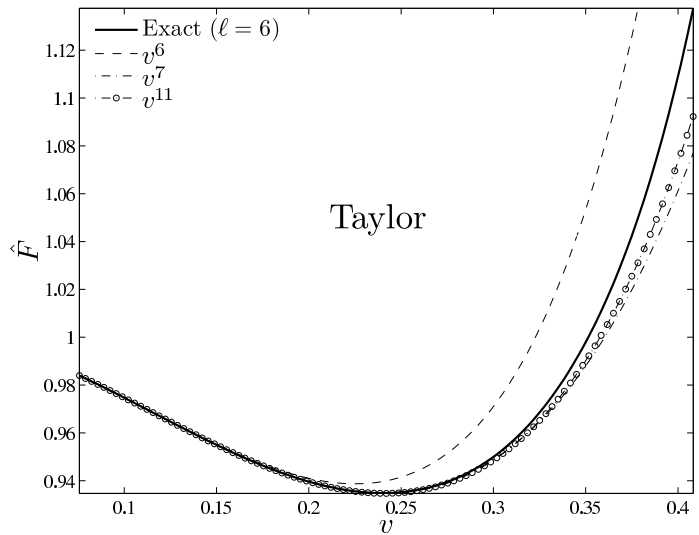}
\caption{The extreme-mass-ratio limit ($\nu=0$): the 
Newton-normalized energy flux emitted by a particle on 
circular orbits. The figure illustrates the 
scattering of the standard Taylor expansion of the flux around
the ``exact'' numerical result (computed up to $\ell=6$) obtained 
via perturbation theory.}
\label{taylor}       
\end{figure}

In the previous Section we have described how the EOB method 
encodes the conservative part of the relative orbital dynamics 
into the dynamics of an 'effective' particle. Let us now 
briefly discuss how to complete the EOB dynamics by defining 
some {\it resummed} expressions describing radiation reaction effects. 
One is interested in circularized binaries, which have lost 
their initial eccentricity under the influence of radiation 
reaction. For such systems, it is enough 
(as shown in~\cite{Buonanno:2000ef}) to include a 
radiation reaction force in the $p_{\varphi}$ equation 
of motion only. More precisely, we are using phase space 
variables $r , p_r , \varphi , p_{\varphi}$ associated 
to polar coordinates (in the equatorial plane 
$\theta = \frac{\pi}{2}$). Actually it is convenient 
to replace the radial momentum $p_r$ by the momentum 
conjugate to the `tortoise' radial 
coordinate $R_* = \int dR (B/A)^{1/2}$, {\it i.e.} $P_{R_*} = (A/B)^{1/2} \,
P_R$. 
The real EOB Hamiltonian is obtained by first solving 
Eq.~(\ref{eqn9}) to get $E_{\rm real}^{\rm total} = \sqrt s$ 
in terms of ${\mathcal E}_{\rm eff}$, and then by solving 
the effective Hamiltonian-Jacobi 
equation\footnote{Completed by the ${\mathcal O} ({\bm p}^4)$ 
terms that must be introduced at 3PN.} to get ${\mathcal E}_{\rm eff}$ 
in terms of the effective phase space coordinates ${\bm q}_{\rm eff}$ 
and ${\bm p}_{\rm eff}$. The result is given by two nested 
square roots (we henceforth set $c=1$):
\begin{equation}
\label{eqn10}
\hat H_{\rm EOB} (r,p_{r_*} , \varphi) = \frac{H_{\rm EOB}^{\rm real}}{\mu} 
= \dfrac{1}{\nu}\sqrt{1 + 2 \nu \, (\hat H_{\rm eff} - 1)} \, ,
\end{equation}
where
\begin{equation}
\label{eqn11}
\hat H_{\rm eff} = \sqrt{p_{r_*}^2 + A(r) \left( 1 + \frac{p_{\varphi}^2}{r^2} + z_3 \, \frac{p_{r_*}^4}{r^2} \right)} \, ,
\end{equation}
with $z_3 = 2\nu \, (4-3\nu)$. Here, we are using suitably rescaled
dimensionless (effective) variables: 
$r = R/GM$, $p_{r_*} = P_{R_*} / \mu$, $p_{\varphi} = P_{\varphi} / \mu \, 
GM$, as well as a rescaled time $t = T/GM$. This leads to equations 
of motion $(r,\varphi,p_{r_*},p_{\varphi})$ of the form
\begin{align}
\frac{d \varphi}{dt}     & =   \frac{\partial \, \hat H_{\rm EOB}}{\partial \, p_{\varphi}}\equiv\Omega \, ,\\
\dfrac{dr}{dt}           & =   \left( \frac{A}{B} \right)^{1/2} \,
\frac{\partial \, \hat H_{\rm EOB}}{\partial \, p_{r_*}} \, ,\\
\label{eqn12}
\frac{d p_{\varphi}}{dt} & =   \hat{\mathcal F}_{\varphi} \,,\\ 
\dfrac{d p_{r_*}}{dt}    & = - \left( \frac{A}{B} \right)^{1/2} \, \frac{\partial \, \hat H_{\rm EOB}}{\partial \, r} \, ,
\end{align}
which explicitly read
\begin{align}
%
\label{eob:1}
\dfrac{d\varphi}{dt}     &= \dfrac{A p_\varphi}{\nu r^2\hat{H}\hat{H}_{\rm eff}} \equiv \Omega\ , \\
\label{eob:2}
\dfrac{dr}{dt}           &= \left(\dfrac{A}{B}\right)^{1/2}\dfrac{1}{\nu\hat{H}\hat{H}_{\rm eff}}\left(p_{r_*}+z_3\dfrac{2A}{r^2}p_{r_*}^3\right) \ , \\
\label{eob:3}
\dfrac{dp_{\varphi}}{dt} &= \hat{\cal F}_{\varphi} \ , \\
\label{eob:4}
\dfrac{dp_{r_*}}{dt}     &= -\left(\dfrac{A}{B}\right)^{1/2}\dfrac{1}{2\nu\hat{H}\hat{H}_{\rm eff}}
  \left\{A'+\dfrac{p_\varphi^2}{r^2}\left(A'-\dfrac{2A}{r}\right)+z_3\left(\dfrac{A'}{r^2}-\dfrac{2A}{r^3}\right)p_{r_*}^4
  \right\} \ ,
\end{align}
where $A'=dA/dr$.
As explained above the EOB metric function $A(r)$ is defined by 
Pad\'e resumming the Taylor-expanded result (\ref{eqn4}) obtained 
from the matching between the real and effective energy levels 
(as we were mentioning, one uses a similar Pad\'e resumming for 
$D(r) \equiv A(r) \, B(r)$). One similarly needs to resum 
${\hat{\cal F}}_{\varphi}$, i.e., the 
$\varphi$ component of the radiation reaction which has been 
introduced on the r.h.s. of Eq.~\eqref{eqn12}. 
During the quasi-circular inspiral $\hat{\mathcal F}_{\varphi}$ 
is known (from the PN work mentioned in Section~2 above) in the 
form of a Taylor expansion of the form
\begin{equation}
\label{eqn13}
\hat{\mathcal F}_{\varphi}^{\rm Taylor} = -\frac{32}{5} \, \nu \, \Omega^5 \, r_{\omega}^4 \, \hat F^{\rm Taylor} (v_{\varphi}) \, ,
\end{equation}
where $v_{\varphi} \equiv \Omega \, r_{\omega}$,
and $r_{\omega}\equiv r[\psi(r,p_\varphi)]^{1/3}$ is a 
modified EOB radius, with $\psi$ being defined as 
\begin{align}
\psi(r,p_\varphi)&=\dfrac{2}{r^2}\left(\dfrac{dA(r)}{dr}\right)^{-1}
                   \left[1+2\nu\left(\sqrt{A(r)\left(1+\dfrac{p_\varphi^2}{r^2}\right)  }-1\right)\right],
\end{align}
which generalizes the 2PN-accurate Eq.~(22) of 
Ref.~\cite{Damour:2006tr}. 
In Eq.~\eqref{eqn13} we have defined
\begin{eqnarray}
\label{eqn14}
\hat F^{\rm Taylor} (v) &= &1 + A_2 (\nu) \, v^2 + A_3 (\nu) \, v^3 + A_4 (\nu) \, v^4 + A_5 (\nu) \, v^5 \nonumber \\
&&+ \, A_6 (\nu , \log v) \, v^6 + A_7 (\nu) \, v^7 + A_8 (\nu = 0 , \log v) \, v^8 \, ,
\end{eqnarray}
where we have added to the known 3.5PN-accurate comparable-mass 
result the small-mass-ratio 4PN contribution~\cite{gr-qc/9405062}.
We recall that the small-mass contribution to the Newton-normalized
flux is actually known up to 5.5PN order, i.e. to $v^{11}$ included.
The standard Taylor expansion of the flux, \eqref{eqn14}, has rather 
poor convergence properties when considered up to the LSO.
This is illustrated
in Fig.~\ref{taylor} in the small-mass limit $\nu=0$.
The convergence of the PN-expanded flux can be studied in detail in the
$\nu=0$ limit, because in this
case one can compute an ``exact'' result numerically (using black
hole perturbation theory~\cite{Cutler:1993vq,Yunes:2008tw}).
The ``exact'' energy flux shown in Fig.~\ref{taylor}
is obtained as a sum over multipoles 
\begin{align}
\label{eq:flux_0}
 F^{\ell max}=\sum_{\ell =2}^{\ell_{\rm max}}\sum_{m=1}^{\ell} F_{\lm},
\end{align}
where $F_{\ell m}=F_{\ell|m|}$ already denotes the sum of two equal 
contributions corresponding to $+m$ and $-m$ ($m\neq0$ as $F_{\ell0}$ 
vanishes for circular orbits).  To be precise, the ``exact'' result 
exhibited in Fig.~\ref{taylor} is given by the rather accurate approximation $F^{(6)}$ 
obtained by choosing $\ell_{\rm max}=6$; i.e., by truncating the sum 
over $\ell$ in Eq.~\eqref{eq:flux_0} beyond $\ell=6$.
In addition, one normalizes the result onto the ``Newtonian'' 
(i.e., quadrupolar) result $F_{22}^N=32/5(\mu/M)^2 v^{10}$. 
In other words, the solid line in Fig.~\ref{taylor} represents the 
quantity $\hat{F}\equiv F^{(6)}/F_{22}^N$.

For clarity, we selected only three Taylor approximants: 
3PN ($v^6$), 3.5PN $(v^7)$ and 5.5PN ($v^{11}$). These three 
values suffice to illustrate the rather large scatter among 
Taylor approximants, and the fact that, near the LSO, the 
convergence towards the exact value (solid line) is rather 
slow, and non monotonic. [See also Fig.~1 in Ref.~\cite{Poisson:1995vs} 
and Fig.~3 of Ref.~\cite{Damour:1997ub} for fuller illustrations 
of the scattered and non monotonic way in which successive 
Taylor expansions approach the numerical result.]
The results shown in Fig.~\ref{taylor} elucidate that the 
Taylor series~\eqref{eqn14} is inadequate to give a reliable 
representation of the energy loss during the
plunge. That is the reason why the EOB formalism advocates 
the use of a ``resummed'' version of ${\cal F}_{\varphi}$, 
i.e. a nonpolynomial function replacing
Eq.~\eqref{eqn14} at the r.h.s. of the Hamilton's equation 
(and coinciding with it in in the $v/c\ll 1$ limit).

Two methods have been proposed to perform such a resummation.
The first method, that strongly relies on the use of 
Pad\'e approximants, was introduced by Damour, 
Iyer and Sathyaprakash~\cite{Damour:1997ub}
and, with different degrees of sophistication, 
has been widely used in the literature dealing 
with the EOB formalism~\cite{Buonanno:2000ef,Buonanno:2005xu,
Buonanno:2006ui,Nagar:2006xv,Pan:2007nw,Buonanno:2007pf,Damour:2007cb,
Damour:2007xr,Damour:2007yf,Damour:2007vq,Damour:2008te,
Boyle:2008ge,Buonanno:2009qa}.
The second resummation method has been recently introduced by Damour, 
Iyer and Nagar~\cite{Damour:2008gu} and exploited to
provide a self-consistent expression of the radiation reaction
force in Ref.~\cite{Damour:2009kr}. This latter resummation 
procedure is based on (i) a new multiplicative decomposition of the 
gravitational metric waveform which yields a (ii) resummation 
of each multipolar contribution to the energy flux.
The use of Pad\'e approximants is a useful tool (but not the 
only one) that proves helpful to further improve the convergence 
properties of each multipolar contribution to the flux. 
The following two Sections are devoted to highlighting the
main features of the two methods. For pedagogical reasons the
calculation is first done in the small-mass limit ($\nu\to 0$)
and then generalized to the comparable mass case.

\subsection{Resummation of $\hat{F}^{\rm Taylor}$ using 
  a one-parameter family of Pad\'e approximants: tuning $v_{\rm pole}$}
\label{sec:vpole}
Following \cite{Damour:1997ub}, one resums $\hat F^{\rm Taylor}$ 
by using the following Pad\'e resummation approach. First, one 
chooses a certain number $v_{\rm pole}$ which is intended to 
represent the value of the orbital velocity $v_{\varphi}$ at 
which the exact angular momentum flux would become infinite 
if one were to formally analytically continue $\hat{\mathcal F}_{\varphi}$ 
along {\it unstable} circular orbits below the Last Stable 
Orbit (LSO): then, given $v_{\rm pole}$, one defines the 
resummed $\hat F (v_{\varphi})$ as
\begin{equation}
\label{eqn15}
\hat F^{\rm resummed} (v_{\varphi}) = \left( 1 - \frac{v_{\varphi}}
{v_{\rm pole}} \right)^{-1} P_4^4 \left[ \left( 1 - \frac{v_{\varphi}}
{v_{\rm pole}} \right) \hat F^{\rm Taylor} (v_{\varphi};\nu=0) \right] \, ,
\end{equation} where $P_4^4$ denotes a $(4,4)$ Pad\'e approximant. 
\begin{figure}[t]
\centering
\includegraphics[height=7 cm]{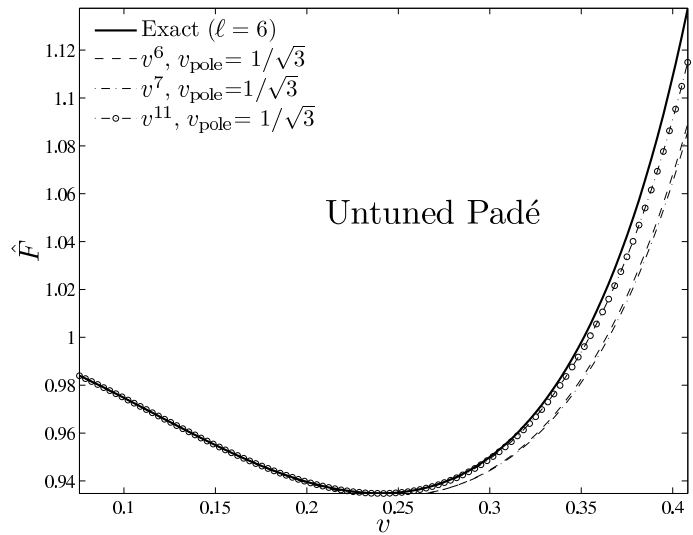}
\caption{The extreme-mass-ratio limit ($\nu=0$). Pad\'e resummation 
of the Taylor expandend energy flux of Fig.~\ref{taylor} as proposed in 
Ref.~\cite{Damour:1997ub} with $v_{\rm pole}=1/\sqrt{3}$.
The sequence of Pad\'e approximants is less scattered than the
corresponding Taylor ones and closer to the exact result.}
\label{pade_untuned}       
\end{figure}

If one first follows the reasoning line of~\cite{Damour:1997ub}, 
and fixes the location of the pole in the 
resummed flux at the standard Schwarzschild 
value $\vp^{(\nu=0)}=1/\sqrt{3}$, one gets
the result in Fig.~\ref{pade_untuned}. 
By comparison to Fig.~\ref{taylor}, one can appreciate the 
significantly better (and monotonic) way in which successive 
{\it Pad\'e approximants} approach (in $L_{\infty}$ norm on the full interval
$0<x<x_{\rm LSO}$) the numerical result. Ref.~\cite{Damour:1997ub} 
also showed that the observationally relevant overlaps (of both the ``faithfulness'' and
the ``effectualness'' types) between analytical and numerical adiabatic
signals were systematically better for Pad\'e approximants than for Taylor
ones. Note that this figure is slightly different from the
corresponding results in panel {\it (b)} of Fig.~3 in~\cite{Damour:1997ub} (in
particular, the present result exhibits a better ``convergence'' of the
$v^{11}$ curve). This difference is due to the new treatment of the
logarithmic terms $\propto \log x$. Instead of factoring them out in front
as proposed in~\cite{Damour:1997ub}, we consider them here
(following~\cite{Damour:2007yf}) as being part of the ``Taylor coefficients''
$f_n(\log x)$ when Pad\'eing the flux function. 
\begin{figure}[t]
\centering
\includegraphics[height=7 cm]{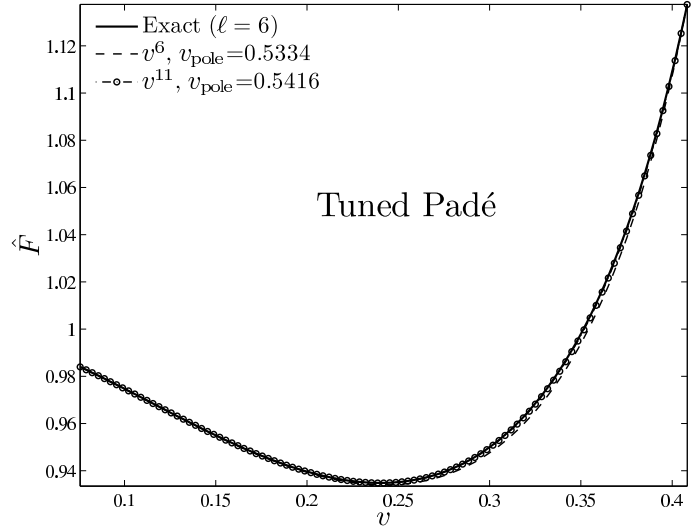}
\caption{The extreme mass ratio limit ($\nu=0$). Same of
 Fig.~\ref{pade_untuned} but {\it flexing} the value of the parameter 
 $v_{\rm pole}$ so to improve the agreement with the exact result.}
\label{tune_vpole}       
\end{figure}

A remarkable 
improvement in the ($L_{\infty}$) closeness between 
$\hat{F}^{\text{Pad\'e-resummed}}(v)$ and $\hat{F}^{\rm Exact}(v)$
can be obtained, as suggested by Damour and Nagar~\cite{Damour:2007yf}
(following ideas originally introduced in Ref.~\cite{Damour:2002vi}),
by suitably flexing the value of $\vp$.
As proposed in Ref.~\cite{Damour:2007yf}, $\vp$ is tuned until
the difference between the resummed and the exact 
flux at the LSO is zero (or at least smaller than $10^{-4}$).
The resulting closeness between the exact and tuned-resummed fluxes 
is illustrated in Fig.~\ref{tune_vpole}.
It is so good (compared to the previous figures, where the differences were
clearly visible) that we need to complement the figure
with Table~\ref{tab:table1}. This table compares in a quantitative way
the result of the ``untuned'' Pad\'e resummation ($\vp=1/\sqrt{3}$) 
of Ref.~\cite{Damour:1997ub} to the result of the ``$\vp$-tuned'' Pad\'e
resummation described here. Defining the function
$\Delta\hat{F}(v;\vp)=\hat{F}^\text{Resummed}(v;\vp)-\hat{F}^\text{Exact}(v)$
measuring the difference between a resummed and the exact energy flux,
Table~\ref{tab:table1} lists both the values of $\Delta\hat{F}$ 
at $v=v_{\rm  LSO}$ and its $L_{\infty}$ norm on the interval $0<v<v_{\rm LSO}$
for both the untuned and tuned cases.
Note, in particular, how the $\vp$-flexing approach permits to reduce 
the $L_{\infty}$ norm over this interval by more than an order of magnitude 
with respect to the untuned case. Note that the closeness between the tuned 
flux and the exact one is remarkably good ($4.3\times 10^{-3}$) already at 
the 3PN level. 
%
%
\begin{table}[t]
\caption{\label{tab:table1} Errors in the flux of the two (untuned or tuned) 
  Pad\'e resummation procedures.  
  From left to right, the columns report: the PN-order; the difference between the
  resummed and the exact flux,  $\Delta\hat{F}=\hat{F}^{\rm
  Resummed}-\hat{F}^{\rm Exact}$, at the LSO,
  and the $L_{\infty}$ norm  of $\Delta\hat{F}$, $||\Delta\hat{F}||_\infty$
  (computed over the interval $0<v<v_{\rm LSO}$), for $\vp=1/\sqrt{3}$;
   the {\it flexed} value of $v_{\rm pole}$ used here;
   $\hat{\Delta}F$ at the LSO and the corresponding $L_{\infty}$ norm 
  (over the same interval) for the flexed value of $\vp$.}
\begin{center}
  \begin{tabular}{ccccccc}
    PN-order   & $\Delta\hat{F}^{1/\sqrt{3}}_{\rm LSO}$ 
               & $||\Delta\hat{F}||_\infty^{1/\sqrt{3}}$
               & $v_{\rm pole}$
               & $\Delta\hat{F}^{\vp}_{\rm LSO}$ 
               & $||\Delta\hat{F}||_\infty^{\vp}$  \\ 
    \hline \hline
 3   ($v^6$)    &   -0.048 &  0.048 & $0.5334$  &  $7.06\times 10^{-5}$ & 0.00426   \\
 3.5 ($v^7$)    &   -0.051 &  0.051 & $0.5425$  &  $5.50\times 10^{-5}$ & 0.00429   \\  
 5.5 ($v^{11}$) &   -0.022 &  0.022 & $0.5416$  &  $2.52\times 10^{-5}$ & 0.000854 
  \end{tabular}
\end{center}
\end{table}%

It has recently been shown in several works~\cite{Damour:2007yf,Damour:2007vq,
Damour:2008te,Buonanno:2009qa} that the {\it flexibility} in the choice of 
$v_{\rm pole}$ could be advantageously used to get a close agreement with NR
data (at the level of the numerical error). We will not comment here any
further on this {\it parameter-dependent} resummation procedure of the 
energy flux and address the reader to the aforementioned references for
further details.

\subsection{Parameter-free resummation of waveform and energy flux}
\label{sec:DIN}

In this section we shall introduce the reader to
the new resummation technique
for the multipolar waveform (and thus for the energy flux) introduced
in Ref.~\cite{Damour:2007xr,Damour:2007yf} and perfected
in~\cite{Damour:2008gu}. The aim is to summarize here the main
ideas discussed in~\cite{Damour:2008gu} as well as to collect 
most of the relevant equations that are useful for implementation in
the EOB dynamics.
To be precise, the new results discussed in Ref.~\cite{Damour:2008gu}
 are twofold: on the one hand, that work generalized
the $\ell=m=2$ resummed waveform of~\cite{Damour:2007xr,Damour:2007yf} 
to higher multipoles by using the most accurate currently known PN-expanded 
results~\cite{Kidder:2007rt,Berti:2007fi,Blanchet:2008je}
as well as the higher PN terms which are known in the test-mass
limit~\cite{Tagoshi:1994sm,Tanaka:1997dj}; on the other hand,
it introduced a {\it new resummation procedure} which consists in 
considering a new theoretical quantity, denoted as $\rho_{\lm}(x)$, 
which enters the $(\ell,m)$ waveform (together with other building 
blocks, see below) only through its $\ell$-th power: $h_{\ell m}\propto
\left(\rho_{\lm}(x)\right)^{\ell}$. Here, and below, $x$ denotes the
invariant PN-ordering parameter $x\equiv (GM\Omega/c^3)^{2/3}$.

The main novelty introduced by Ref.~\cite{Damour:2008gu} is 
to write the $(\ell,m)$ multipolar waveform emitted by a 
circular nonspinning compact binary as the {\it product} of several 
factors, namely
\begin{align}
\label{eq:hlm}
h_{\lm}^{(\epsilon)}&=\dfrac{GM\nu}{c^2 R} n_{\lm}^{(\epsilon)} c_{\l+\epsilon}(\nu)
x^{(\ell+\epsilon)/2}Y^{\ell-\epsilon,-m}\left(\dfrac{\pi}{2},\Phi\right)
                  \hat{S}_{\rm  eff}^{(\epsilon)}T_{\ell m} e^{\ii\delta_{\lm}} \rho_{\lm}^\ell.
\end{align}
Here $\epsilon$ denotes the parity of $\ell+m$ ($\epsilon=\pi(\ell+m)$), i.e.
$\epsilon=0$ for ``even-parity'' (mass-generated) multipoles ($\ell+m$ even), and
$\epsilon=1$ for ``odd-parity'' (current-generated) ones ($\ell+m$ odd); $n_{\lm}^{(\epsilon)}$ and
$c_{\l+\epsilon}(\nu)$ are numerical coefficients;
$\hat{S}^{(\epsilon)}_{\rm eff}$ is a $\mu$-normalized effective source 
(whose definition comes from the EOB formalism); $T_{\ell m}$ is a resummed
version~\cite{Damour:2007xr,Damour:2007yf} of an infinite number of
``leading logarithms'' entering the {\it tail effects}~\cite{Blanchet:1992br,Blanchet:1997jj};
$\delta_{\lm}$ is a supplementary phase (which corrects the phase effects not
included in the {\it complex} tail factor $T_{\lm}$), and, finally,
$\left(\rho_{\lm}\right)^\ell$ denotes the $\ell$-th power of the quantity
$\rho_{\lm}$ which is the new building block introduced
in~\cite{Damour:2008gu}. Note that in previous
papers~\cite{Damour:2007xr,Damour:2007yf}  the quantity 
$\left(\rho_{\lm}\right)^\ell$ was denoted as $f_{\lm}$
and we will mainly use this notation below.
Before introducing explicitly the various elements entering the 
waveform \eqref{eq:hlm} it is convenient to 
decompose $h_{\lm}$ as 
\begin{equation}
\label{hlm_expanded}
h_{\lm} = h_{\lm}^{(N,\epsilon)} \hat{h}_{\lm}^{(\epsilon)},
\end{equation}
where $h_{\lm}^{(N,\epsilon)}$ is the Newtonian contribution 
and $\hat{h}_{\lm}^{(\epsilon)}\equiv \hat{S}_{\rm eff}^{(\epsilon)}
T_{\lm}e^{\rm i\delta_{\lm}}f_{\lm}$ represents a resummed
version of all the PN corrections. The PN correcting factor 
$\hat{h}_{\lm}^{(\epsilon)}$, as well as all its building blocks, 
has the structure
$\hat{h^{(\epsilon)}}_{\lm}=1+{\cal O}(x)$.

Entering now in the discussion of the explicit form of the elements 
entering Eq.~\eqref{eq:hlm}, we have that the $\nu$-independent
numerical coefficients are given by
\begin{align}
\label{eq:newtnorm}
n^{(0)}_{\lm}  & = (\ii m)^{\ell} \dfrac{8\pi}{(2\ell +
  1)!!}\sqrt{\dfrac{(\ell+1)(\ell+2)}{\ell(\ell-1)}}, \\
n^{(1)}_{\lm} & = -(\ii m)^\ell \dfrac{16\pi\ii}{(2\ell+1)!!}\sqrt{\dfrac{(2\ell+1)(\ell+2)(\ell^2-m^2)}{(2\ell-1)(\ell+1)\ell(\ell-1)} },
\end{align}
while the $\nu$-dependent coefficients 
$c_{\ell+\epsilon}(\nu)$ (such that
\hbox{$|c_{\ell+\epsilon}(\nu=0)|=1$}), 
can be expressed in terms of $\nu$ (as in Ref.~\cite{Kidder:2007rt,Blanchet:2008je}), 
although they are more conveniently
written in terms of the two mass ratios $X_1=m_1/M$ and $X_2=m_2/M$ 
in the form
\begin{align}
\label{eq:cl}
c_{\ell+\epsilon}(\nu) & 
= X_2^{\ell+\epsilon-1}+(-)^{\ell+\epsilon}X_1^{\ell+\epsilon-1}\nonumber\\
&=X_2^{\ell+\epsilon-1}+(-)^m X_1^{\ell+\epsilon-1}.
\end{align}
In the second form of the equation we have used the fact that, 
as $\epsilon=\pi(\ell+m)$, $\pi(\ell+\epsilon)=\pi(m)$.

Let us turn now to discussing the structure of the $\hat{S}^{(\epsilon)}_{\rm eff}$
and $T_{\lm}$ factors. To this aim, following Ref.~\cite{Damour:2008gu}, we
recall that the along the sequence of EOB circular orbits, which are 
determined by the condition $\de_u\left\{A(u)[1+ j_0^2\, u^2]\right\}=0$,
the effective EOB Hamiltonian (per unit $\mu$ mass) reads
\begin{equation}
\label{eq:Heff}
\hat{H}_{\rm eff}=\dfrac{H_{\rm eff}}{\mu}=\sqrt{A(u)(1+j_0^2\, u^2)} \quad
\text{(circular orbits)}.
\end{equation}
where the squared angular momentum is given by
\begin{equation}
\label{eq:j0}
j_0^2(u)=-\dfrac{A'(u)}{(u^2 A(u))'}\quad\text{(circular orbits)},
\end{equation}
with the prime denoting  $d/du$. Inserting this $u$-parametric representation of 
$j^2$ in Eq.~\eqref{eq:Heff} defines the $u$-parametric representation of the
effective Hamiltonian $\hat{H}_{\rm eff}(u)$.
In the even-parity case (corresponding to mass
moments), since  the leading order source of gravitational radiation 
is given by the energy density, Ref.~\cite{Damour:2008gu} defined
the even-parity ``source factor'' as
\begin{align}
\label{eq:source_even}
\hat{S}^{(0)}_{\rm eff}(x) &=\hat{H}_{\rm eff}(x)\qquad \ell+m \quad \text{even},
\end{align}
where $x=(GM\Omega/c^3)^{2/3}$.
In the odd-parity case, they explored two, equally motivated, possibilities. 
The first one consists simply in still factoring $\hat{H}_{\rm eff}(x)$;
i.e., in defining 
$\hat{S}^{(1,H)}_{\rm eff} =\hat{H}_{\rm eff}(x)$ also when 
$\ell+m$ is odd.
The second one consists in factoring the angular momentum ${\cal J}$.
Indeed, the angular momentum density $\epsilon_{ijk}x^j \tau^{0k}$
enters as a factor in the (odd-parity) current moments, and ${\cal J}$ 
occurs (in the small-$\nu$ limit) as a 
factor in the source of the Regge-Wheeler-Zerilli odd-parity multipoles.
This leads us to define as second possibility
\begin{equation}
\label{eq:source_odd_J}
\hat{S}^{(1,J)}_{\rm eff} =\hat{j}(x)\equiv x^{1/2}j(x) \qquad  \ell+m \quad \text{odd},
\end{equation}
where $\hat{j}$ denotes what can be called the ``Newton-normalized'' 
angular momentum, namely the ratio $\hat{j}(x)=j(x)/j_N(x)$
with $j_N(x)=1/\sqrt{x}$. 
In Ref.~\cite{Damour:2008gu} the relative merits of the two possible choices
were discussed. Although the analysis in the adiabatic $\nu=0$ limit showed 
that they are equivalent from the practical point of view (because they both 
yield waveforms that are very close to the exact numerical result) we prefer
to consider only the $J$-factorization in the following, that we will
treat as our standard choice.

The second building block in our factorized decomposition is the ``tail
factor'' $T_{\lm}$ (introduced in Refs.~\cite{Damour:2007xr,Damour:2007yf}).
As mentioned above, $T_{\lm}$ is a resummed version of an infinite number 
of ``leading logarithms'' entering the transfer function between the 
near-zone multipolar wave and the far-zone one, 
due to {\it tail effects} linked to its propagation in a Schwarzschild 
background of mass $M_{\rm ADM}=H^{\rm real}_{\rm EOB}$. 
Its explicit expression reads
\begin{equation}
\label{eq:tail_factor}
T_{\lm} = \dfrac{\Gamma(\ell+1-2\ii\hat{\hat{k}})}{\Gamma(\ell
  +1)}e^{\pi\hat{\hat{k}}}e^{2\ii\hat{\hat{k}}\log(2 k r_0)} ,
\end{equation}
where $r_0=2GM$ and $\k\equiv G H^{\rm real}_{\rm EOB} m\Omega$
and $k\equiv m\Omega$.
Note that  $\k$ differs from $k$ by a rescaling involving 
the {\it real} (rather than the {\it effective}) 
EOB Hamiltonian, computed at this stage along the sequence of
circular orbits.

The tail factor $T_{\lm}$ is a complex number which already takes into 
account some of the dephasing of the partial waves as they propagate
out from the near zone to infinity. However, as the tail factor only takes 
into account the leading logarithms, one needs to correct it by a complementary 
dephasing term, $e^{\ii\delta_{\lm}}$,  
linked to subleading logarithms and other effects.
This subleading phase correction can be computed as being the phase
$\delta_{\lm}$ of the
complex ratio between the PN-expanded $\hat{h}_{\lm}^{(\epsilon)}$ and the 
above defined source and tail factors. In the comparable-mass case
($\nu\neq0$), the 3PN $\delta_{22}$ phase correction to the leading quadrupolar
wave was originally computed in Ref.~\cite{Damour:2007yf} (see also
Ref.~\cite{Damour:2007xr} for the $\nu=0$ limit). Full results for
the subleading partial waves to the highest  
possible PN-accuracy by starting from the currently known 
3PN-accurate $\nu$-dependent waveform~\cite{Blanchet:2008je}
have been obtained in~\cite{Damour:2008gu}.

The last factor in the multiplicative decomposition
of the multipolar waveform can be computed 
as being the modulus $f_{\lm}$ of the complex ratio between 
the PN-expanded $\hat{h}_{\lm}^{(\epsilon)}$  and the 
above defined source and tail factors.
In the comparable mass case
($\nu\neq0$), the $f_{22}$ modulus correction to the leading quadrupolar
wave was computed in Ref.~\cite{Damour:2007yf} (see also
Ref.~\cite{Damour:2007xr} for the $\nu=0$ limit). 
For the  subleading partial waves, Ref.~\cite{Damour:2008gu}
explicitly computed the other $f_{\lm}$'s to the highest 
possible PN-accuracy by starting from the currently known 
3PN-accurate $\nu$-dependent waveform~\cite{Blanchet:2008je}.
In addition, as originally proposed in Ref.~\cite{Damour:2007yf}, 
to reach greater accuracy the $f_{\lm}(x;\nu)$'s extracted from
the 3PN-accurate $\nu\neq 0$ results  are completed by adding 
higher order contributions coming from the 
$\nu=0$ results~\cite{Tagoshi:1994sm,Tanaka:1997dj}.
In the particular $f_{22}$ case discussed 
in~\cite{Damour:2007yf}, this amounted to adding 4PN and 5PN $\nu=0$
terms. This ``hybridization'' procedure was then systematically 
pursued for all the other multipoles, using the 5.5PN accurate 
calculation of the multipolar decomposition of the gravitational 
wave energy flux of Refs.~\cite{Tagoshi:1994sm,Tanaka:1997dj}.
Note that such hybridization procedure 
is {\it not} equivalent to the
straightforward hybrid sum ansatz, 
$\tilde{h}_{\ell m}=\tilde{h}_{\ell m}^{\text{known}}(\nu)+\tilde{h}_{\ell
  m}^{\text{higher}}(\nu=0)$ (where $\tilde{h}_{\lm}\equiv h_{\lm}/\nu$)
that one may have thought to implement.

In the even-parity case, the determination of the modulus $f_{\lm}$ is unique. 
In the odd-parity case, it depends on the choice of the source which, as
explained above, can be connected either to the effective energy  or 
to the angular momentum. We will consider both cases and 
distinguish them by adding either the label $H$ or $\J$ to
the corresponding $f_{\lm}$.  Note, in passing, that, since 
in both cases the factorized effective source term ($H_{\rm eff}$ or $\J$) 
is a real quantity, the phases $\delta_{\lm}$'s are the same.

The above explained procedure defines the $f_{\lm}$'s as Taylor-expanded
PN series of the type
\begin{equation}
\label{scheme_flm}
f_{\lm}(x;\nu) = 1 + c_1^{f_{\lm}}(\nu) x + c_2^{f_\lm}(\nu) x^2 + c_3^{f_\lm}(\nu, \log(x))x^3 + \dots
\end{equation}
Note that one of the virtues of our factorization is to
have separated the half-integer powers of $x$ appearing in the usual 
PN-expansion of $h_{\lm}^{(\epsilon)}$ from the integer powers, the tail
factor, together with the complementary phase factor $e^{\ii\delta_{\lm}}$,
having absorbed all the half-integer powers.
In Ref.~\cite{Damour:2008te} all the $f_{\ell m}$'s (both for the $H$ and $\J$
choices) have been computed up to the highest available ($\nu$-dependent or
not) PN accuracy. In the formulas for the $f_{\lm}$'s given below we
``hybridize'' them by adding to the known $\nu$-dependent coefficients $c_{n}^{f_{\lm}}(\nu)$
in Eq.~\eqref{scheme_flm} the $\nu=0$ value of the higher order coefficients:
$c_{n'}^{f_{\lm}}(\nu=0)$.
The 1PN-accurate $f_{\lm}$'s for  $\ell+m$ even 
and and also for $\ell+m$ odd  can be written down for all $\ell$.
The complete result for the $f_{\lm}$'s
that are known with an accuracy higher than 1PN
are listed in Appendix~B of Ref.~\cite{Damour:2008te}.
Here, for illustrative purposes, we quote only the 
lowest  $f_\lm^{\text{even}}$ and $f_{\lm}^{\text{odd},J}$ 
up to $\ell=3$ included.
\begin{align}
\label{eq:f22}
&f_{22}(x;\nu) = 1 +\frac{1}{42} (55 \nu -86) x + 
\frac{\left(2047 \nu ^2-6745 \nu -4288\right)}{1512} x^2 \nonumber \\
  &+\left(\frac{114635 \nu ^3}{99792}-\frac{227875 \nu ^2}{33264}+
   \frac{41}{96}\pi^2\nu-\frac{34625 \nu }{3696}-\frac{856}{105}
   \text{eulerlog}_{2}(x)+\frac{21428357}{727650}\right) x^3\nonumber\\ 
  &+\left(\frac{36808}{2205}\text{eulerlog}_{2}(x)-\frac{5391582359}{198648450}\right)
  x^4\nonumber\\
&+\left(\frac{458816}{19845}\text{eulerlog}_{2}(x)-\frac{93684531406}{893918025}\right)x^5
+{\cal O}(x^6),
\end{align}
\begin{align}
&f^J_{21}(x;\nu)=1+\left(\frac{23 \nu}{42}-\frac{59}{28}\right) x
  +\left(\frac{85 \nu ^2}{252}-\frac{269\nu }{126}-\frac{5}{9}\right)
  x^2\nonumber\\
&+\left(\frac{88404893}{11642400}-\frac{214}{105}\text{eulerlog}_1(x)\right) x^3 \nonumber\\
&+\left(\frac{6313}{1470}\text{eulerlog}_1(x)-\frac{33998136553}{4237833600}\right)x^4+{\cal
  O}(x^5),
\end{align}
\begin{align}
&f_{33}(x;\nu)= 1 +\left(2 \nu -\frac{7}{2}\right) x
+\left(\frac{887 \nu ^2}{330}-\frac{3401 \nu}{330}-\frac{443}{440}\right) x^2 \nonumber\\
&+\left(\frac{147471561}{2802800}-\frac{78}{7} \text{eulerlog}_{3}(x)\right)x^3
+\left(39\;\text{eulerlog}_{3}(x)-\frac{53641811}{457600}\right) x^4+{\cal O}(x^5), 
\end{align}
\begin{align}
&f^J_{32}(x;\nu) = 1+\frac{320 \nu ^2-1115 \nu +328}{90 (3 \nu -1)}x
+\frac{39544 \nu ^3-253768 \nu ^2+117215 \nu -20496}{11880(3\nu -1)}x^2\nonumber\\
&+\left(\frac{110842222}{4729725}-\frac{104}{21}\text{eulerlog}_{2}(x)\right)
x^3+{\cal O}(x^4),
\end{align}
\begin{align}
\label{eq:f31}
&f_{31}(x;\nu) = 1 + \left(-\frac{2 \nu }{3}-\frac{13}{6}\right)x+\left(-\frac{247 \nu ^2}{198}-\frac{371 \nu }{198}+\frac{1273}{792}\right) x^2\nonumber\\
              & +\left(\frac{400427563}{75675600}-\frac{26}{21}\text{eulerlog}_{1}(x)\right)x^3
               +\left(\frac{169}{63}
              \text{eulerlog}_{1}(x)-\frac{12064573043}{1816214400}\right)
              x^4+{\cal O}(x^5). 
\end{align}
For convenience and readability, we have introduced the 
following ``eulerlog'' functions $\text{eulerlog}_{m}(x)$
$\text{eulerlog}_m(x)  = \gamma_E + \log 2+\dfrac{1}{2}\log x + \log m$,
where \hbox{$\gamma_E=0.57721\dots$} is Euler's constant.

The decomposition of the total PN-correction factor
$\hat{h}_{\lm}^{(\epsilon)}$
into several factors is in itself a resummation procedure which has already
improved the convergence of the PN series one has to deal with:
indeed, one can see that the coefficients entering increasing powers of $x$ in the
$f_{\lm}$'s tend to be systematically smaller than the coefficients appearing
in the usual PN expansion of $\hat{h}_{\lm}^{(\epsilon)}$. The reason for this
is essentially twofold: (i) the factorization of $T_{\lm}$ has absorbed powers 
of $m\pi$ which contributed to make large coefficients in
$\hat{h}_{\lm}^{(\epsilon)}$,
and (ii) the factorization of either $\hat{H}_{\rm eff}$ or $\hat{j}$ has
(in the $\nu=0$ case)  removed the presence of an inverse square-root singularity
located at $x=1/3$ 
which caused the coefficient of $x^n$ in any PN-expanded quantity to grow
as $3^{n}$ as $n\to\infty$.
To prevent some potential misunderstandings, let us emphasize that we are
talking here about a singularity entering the analytic continuation (to
larger values of $x$) of a mathematical function $h(x)$ defined (for small
values of $x$) by considering the formal adiabatic circular limit. The point
is that, in the $\nu\to 0$ limit, the radius of convergence and therefore the
growth with $n$ of the PN coefficients of $h(x)$ (Taylor-expanded at $x=0$),
are linked to the singularity of the analytically continued $h(x)$ which is
nearest to $x=0$ in the complex $x$-plane.
In the $\nu\to 0$ case, the nearest singularity in the complex $x$-plane comes
from the source factor $\hat{H}_{\rm eff}(x)$ or $\hat{j}(x)$ in the waveform
and is located at the light-ring $x_{\rm LR}(\nu=0)=1/3$. In the $\nu\neq 0$
case, the EOB formalism transforms the latter (inverse square-root)
singularity in a more complicated (``branching'') singularity where
$d\hat{H}_{\rm eff}/dx$ and $d\hat{j}/dx$ have inverse square-root
singularities located at what 
is called~\cite{Buonanno:2000ef,Buonanno:2006ui,Buonanno:2007pf,Damour:2007vq,Damour:2007yf} 
the (Effective)\footnote{Beware that this ``Effective EOB-light-ring'' occurs
for a circular-orbit radius slightly larger than the purely dynamical
(circular) EOB-light-ring (where $H_{\rm eff}$ and ${\cal J}$ 
would formally become infinite). } 
``EOB-light-ring'', i.e., the (adiabatic) maximum of $\Omega$,
$x_{\rm ELR}^{\rm adiab}(\nu)\equiv 
\left(M\Omega^{\text{adiab}}_{\rm max}\right)^{2/3}\gtrsim 1/3$.

Despite this improvement, the resulting ``convergence'' of the usual
Taylor-expanded $f_{\lm}(x)$'s quoted above does not seem to be good 
enough, especially near or below the LSO, in view of the 
high-accuracy needed to define gravitational wave templates.
For this reason, Refs.~\cite{Damour:2007xr,Damour:2007yf} proposed
to further resum the $f_{22}(x)$ function via a Pad\'e (3,2) approximant,
$P^3_{2}\{f_{22}(x;\nu)\}$, so as to improve its behavior in the
strong-field-fast-motion regime. Such a resummation gave an excellent
agreement with numerically computed waveforms, near the end of the inspiral 
and during the beginning of the plunge, 
for different mass ratios~\cite{Damour:2007xr,Damour:2007vq,Damour:2008te}.
As we were mentioning above, a new route for resumming $f_{\lm}$
was explored in Ref.~\cite{Damour:2008gu}. It is 
based on replacing $f_{\lm}$ by its $\ell$-th root, say
\begin{equation}
\label{eq:lth_root}
\rho_{\lm}(x;\nu) = [f_{\lm}(x;\nu)]^{1/\ell}.
\end{equation}
The basic motivation for replacing $f_{\ell m}$ by $\rho_{\ell m}$ 
is the following: the leading ``Newtonian-level'' contribution 
to the waveform $h^{(\epsilon)}_{\ell m}$ contains a  factor 
$\omega^\ell r_{\rm harm}^\ell v^\epsilon$ where $r_{\rm harm}$  is the
harmonic radial coordinate used in the MPM 
formalism~\cite{Blanchet:1989ki,Damour:1990ji} .
When computing the PN expansion of this factor one has to insert 
the PN expansion of the (dimensionless) harmonic radial 
coordinate $r_{\rm harm}$, $ r_{\rm harm} = x^{-1}(1+c_1 x+{\cal O }(x^2))$,
as a function of the gauge-independent
frequency parameter $x$. 
The PN re-expansion of $[r_{\rm harm}(x)]^\ell$ then generates terms of the 
type $x^{-\ell}(1 +\ell c_1 x+....)$. 
This is one (though not the only one) of the origins of 
1PN corrections in $h_{\ell m}$ and $f_{\ell m}$ 
whose coefficients grow linearly with $\ell$.
The study of~\cite{Damour:2008gu} has pointed out that
these $\ell$-growing terms are problematic for
the accuracy of the PN-expansions. 
Our replacement of $f_{\ell m}$ by $\rho_{\ell m}$ 
is a cure for this problem.
More explicitly, the 
the investigation of $1$PN corrections to 
GW amplitudes~\cite{Blanchet:1989ki,Damour:1990ji,Kidder:2007rt} 
has shown that, in the even-parity case 
(but see also Appendix~A of Ref.~\cite{Damour:2008gu} 
for the odd-parity case), 
\begin{align}
\label{c1lm_nu}
 c_1^{f_{\lm}}(\nu)= 
-\ell\left(1-\dfrac{\nu}{3}\right)+\dfrac{1}{2}
+\dfrac{3}{2}\dfrac{c_{\ell+2}(\nu)}{c_\ell(\nu)}
-\dfrac{b_\ell(\nu)}{c_\ell(\nu)}-\dfrac{c_{\ell+2}(\nu)}{c_{\ell}(\nu)}\dfrac{m^2(\ell+9)}{2(\ell+1)(2\ell+3)},
\end{align}
where $c_{\ell}(\nu)$ is defined in Eq.~\eqref{eq:cl} and
\begin{equation}
b_{\ell}(\nu)\equiv X^{\ell}_2+(-)^{\ell}X_1^{\ell}.
\label{eq:bl}
\end{equation}
Focusing on the $\nu=0$ case for simplicitly 
(since the $\nu$ dependence of $c_1^{f_{\lm}}(\nu)$ is quite
mild~\cite{Damour:2008gu}),
the above result shows that the PN expansion 
of $f_{\lm}$ starts as
\begin{equation}
\label{flm_1PN_even}
f_{\lm}^{\rm even}(x;0)  = 1-\ell x \left(1-\dfrac{1}{\ell}
+\dfrac{m^2(\ell+9)}{2\ell(\ell+1)(2\ell+3)}\right) + {\cal O}(x^2) .
\end{equation}
The crucial thing to note in this result is that as $\ell$ gets large (keeping
in mind that $|m|\leq \ell$), the coefficient of $x$ will be negative and will
approximately range between $-5\ell/4$ and $-\ell$. This means that when 
$\ell \geq 6$ the 1PN correction in $f_{\lm}$ would by itself make
$f_{\lm}(x)$ vanish before the ($\nu=0$) LSO $x_{\rm LSO}=1/6$.
For example, for the 
$\ell=m=6$ mode, one has $f^{\rm 1PN}_{66}(x;0)=1-6x(1+11/42)\approx 1 - 6x(1+ 0.26)$
which means a correction equal to $-100\%$ at $x=1/7.57$ and
larger than $-100\%$ at the LSO, namely $f^{\rm 1PN}_{66}(1/6;0)\approx 1 - 1.26=-0.26$.
This value is totally incompatible with the ``exact'' value 
$f_{22}^{\rm exact}(x_{\rm LSO})=0.66314511$ computed from numerical 
data in Ref.~\cite{Damour:2008gu}.
\begin{figure}[t]
\centering
\includegraphics[height=7 cm]{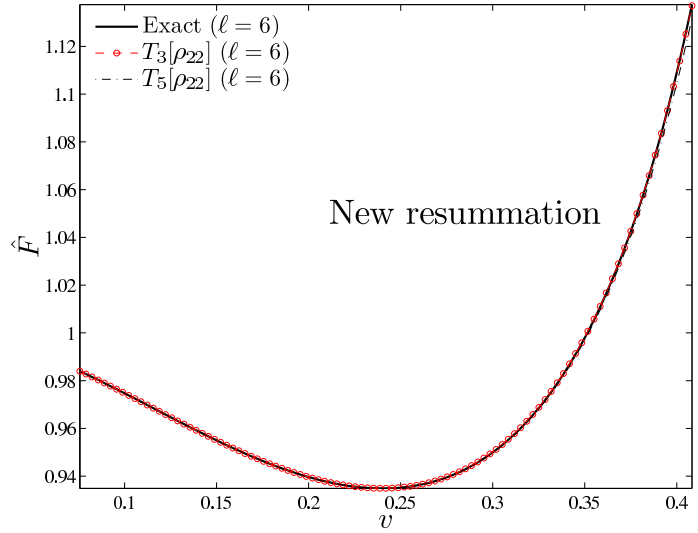}
\caption{Performance of the new resummation procedure described  
in Ref.~\cite{Damour:2008gu}. The total GW flux $\hat{F}$ (up to 
$\ell_{\rm max}=6$) computed from inserting in Eq.~\eqref{eq:flux_1}
the factorized waveform~\eqref{eq:hlm} with the Taylor-expanded
$\rho_{\lm}$'s (with either 3PN or 5PN accuracy for $\rho_{22}$)
is compared with the ``exact'' numerical data.
\label{new_flux} }
\end{figure}

Finally, one uses the newly resummed multipolar waveforms~\eqref{eq:hlm}
to define a resummation of the {\it radiation reaction force} 
$\cal F_{\varphi}$ is defined as
\begin{equation}
\label{eq:RR_new}
{\cal F}_{\varphi} = -\dfrac{1}{\Omega} F^{(\ell_{\rm max})},
\end{equation}
where the (instantaneous, circular) GW flux $F^{(\ell_{\rm max})}$ is defined 
as 
\begin{align}
\label{eq:flux_1}
 F^{(\ell_{\rm max})}=\dfrac{2}{16\pi G}\sum_{\ell =2}^{\ell_{\rm
 max}}\sum_{m=1}^{\ell}|R\dot{h}_{\lm}|^2
=\dfrac{2}{16\pi G}\sum_{\ell =2}^{\ell_{\rm max}}\sum_{m=1}^{\ell}(m\Omega)^2|R h_{\lm}|^2.
\end{align}
As an example of the performance of the new resummation procedure
based on the decomposition of $h_{\lm}$ given by Eq.~\eqref{eq:hlm},
let us focus, as before, on the computation of the GW energy flux
emitted by a test particle on circular orbits on Schwarzschild spacetime.
Figure~\ref{new_flux} illustrates the 
remarkable improvement in the closeness between
$\hat{F}^{\text{New-resummed}}$ and $\hat{F}^{\rm Exact}$.
The reader should compare this result with the previous 
Fig.~\ref{taylor} (the straightforward Taylor approximants to the flux),
Fig.~\ref{pade_untuned} (the Pad\'e resummation with $v_{\rm pole}=1/\sqrt{3}$)
and Fig.~\ref{tune_vpole} (the $v_{\rm pole}$-tuned Pad\'e resummation).
To be fully precise, Fig.~\ref{new_flux} plots two examples of 
fluxes obtained from our new $\rho_{\lm}$-representation for 
the individual multipolar waveforms $h_{\lm}$.
These two examples differ in the choice of approximants for the $\ell=m=2$
partial wave. One example uses for $\rho_{22}$ its 3PN Taylor expansion,
$T_3[\rho_{22}]$, while the other one uses its 5PN Taylor expansion,
$T_5[\rho_{22}]$. All the other partial waves are given by their maximum known
Taylor expansion\footnote{We recall that Ref.~\cite{Damour:2008gu} has also 
shown that the agreement improves even more when the Taylor expansion of the
function $\rho_{22}$ is further suitably Pad\'e resummed.}.
Note that the fact that we use here for the $\rho_{\lm}$'s
some straightforward Taylor expansions does not mean that this new procedure 
is not a resummation technique. Indeed, the defining resummation features of
our procedure have four sources: (i) the factorization of the PN
corrections to the waveforms into four different blocks, namely 
$\hat{S}_{\rm  eff}^{(\epsilon)}$, $T_{\lm}$, $e^{\ii \delta_{\lm}}$ and
$\rho_{\lm}^{\ell}$ in Eq.~\eqref{eq:hlm}; (ii) the fact 
the $\hat{S}_{\rm  eff}^{(\epsilon)}$ is by itself a resummed source whose
PN expansion would contain an infinite number of terms; (iii) the fact that
the tail factor is a closed form expression (see Eq.~\eqref{eq:tail_factor}
above) whose PN expansion also contains an infinite number of terms and 
(iv) the fact that we have replaced the Taylor expansion of 
$f_{\lm}\equiv \rho_{\lm}^\ell$ by that of its $\ell$-th root, namely $\rho_{\lm}$.

In conclusion, Eqs.~\eqref{eq:hlm} and \eqref{eq:flux_1} introduce a new
recipe to resum the ($\nu$-dependent) GW energy flux that is alternative to
the ($v_{\rm pole}$-tuned) one given by Eq.~\eqref{eqn15}.
The two main advantages of the new resummation are: (i) it gives a better
representation of the exact result in the $\nu\to 0$ limit (compare
Fig.~\ref{new_flux} to Fig.~\ref{tune_vpole}), and (ii) it is {\it parameter-free}:
the only flexibility that one has in the definition of the waveform
and flux is the choice of the analytical representation of the function $f_{22}$, 
like, for instance, $P^{3}_{2}\left\{f_{22}\right\}$, 
$\left(T_3\left[\rho_{22}\right]\right)^2$, 
$\left(T_5\left[\rho_{22}\right]\right)^2$, etc., 
(although Ref.~\cite{Damour:2008gu} has pointed out the good consistency 
among all these choices).
Note, that when $\nu\neq 0$, the GW energy flux will 
depend on the choice of resummation of the radial potential $A(R)$
through the Hamiltonian (for the even-parity modes) or the 
angular momentum (for the odd-parity modes).
At the practical level, this means that the EOB model, implemented 
with the new resummation procedure of the energy flux (and waveform) described
so far, will essentially only depend on the doublet of parameters 
$(a_5,a_6)$, that can in principle be constrained by comparison with
(accurate) numerical relativity results.
Contrary to the previous $v_{\rm pole}$-resummation of the radiation
reaction, this route to resummation is free of radiation-reaction 
flexibility parameters.
We will consider it as our ``standard'' route to the resummation of 
the energy flux in the following Sections discussing in details the
properties of the EOB dynamics and waveforms.

\section{Effective One Body dynamics and waveforms}
\label{sec:5}

In this section we marry together all the EOB building blocks described
in the previous Sections and discuss the characteristic of the dynamics
of the two black holes as provided by the EOB approach.
In the following three subsections we discuss in some detail: 
(i) the set up of initial data for the EOB dynamics  with 
negligible eccentricity (Sec.~\ref{ppc:id}); (ii) the structure of
the full Effective One Body waveform, covering inspiral, plunge, merger
and ringdown, with the introduction of suitable Next-to-Quasi-Circular (NQC)
effective corrections to it (and thus to the energy
flux) (Sec.~\ref{waves}); (iii) the explicit structure of the EOB dynamics,
discussing the solution of the dynamical equations.

\subsection{Post-post-circular initial data}
\label{ppc:id}

In this section we discuss in detail the so-called {\it post-post-circular}
 dynamical initial data (positions and momenta) as 
introduced in Sec.~III~B of~\cite{Damour:2007yf}.
This kind of (improved) construction is needed to have
initial data with negligible eccentricity. Since the construction
of the initial data is analytical, including the correction is
useful to start the system relatively close and to avoid evolving
the EOB equation of motion for a long time in order to make
the system circularize itself.

To explain the improved construction of initial
data let us introduce a formal book-keeping parameter $\varepsilon$
(to be set to 1 at the end) in front of the radiation reaction 
$\hat{\cal F}_\varphi$ in the EOB equations of motion. One
can then show that the quasi-circular inspiralling solution of 
the EOB equations of motion formally satisfies
\begin{align}
p_{\varphi}& = j_0(r) + \varepsilon^2 j_2(r) + O(\varepsilon^4), \\
p_{r_*}    & =\varepsilon\pi_1(r) + \varepsilon^3\pi_3(r) + O(\varepsilon^5).
\end{align}
Here, $j_0(r)$ is the usual {\it circular} approximation to
the inspiralling angular momentum as explicitly given 
by Eq.~\eqref{eq:j0} above.
The order $\varepsilon$ 
(``post-circular'') term $\pi_1(r)$ is obtained by: 
(i) inserting the circular approximation 
$p_\varphi=j_0(r)$ on the left-hand side (l.h.s) of Eq.~(10)
of~\cite{Damour:2007cb}, (ii) using the chain rule
$dj_0(r)/dt=(dj_0(r)/dr)(dr/dt)$, (iii) replacing $dr/dt$ by
the right-hand side (r.h.s) of Eq.~(9) of~\cite{Damour:2007cb}
and (iv) solving for $p_{r_*}$ at the first order in $\varepsilon$.
This leads to an explicit result of the form (using the notation
defined in Ref.~\cite{Damour:2007cb})
\begin{align}
\label{p_adiab}
\varepsilon \pi_1(r)
= \left[\nu\hat{H}\hat{H}_{\rm eff}\left(\dfrac{B}{A}\right)^{1/2}\left(\dfrac{dj_0}{dr}\right)^{-1}\hat{\cal
    F}_\varphi\right]_0,
\end{align} 
where the subscript $0$ indicates that the r.h.s. is evaluated
at the leading circular approximation $\varepsilon\to 0$.
The post-circular EOB approximation $(j_0,\pi_1)$ was introduced
in Ref.~\cite{Buonanno:2000ef} and then used in most of the subsequent
EOB papers~\cite{Buonanno:2005xu,Buonanno:2006ui,
Pan:2007nw,Buonanno:2007pf,Damour:2007cb,Nagar:2006xv}.
The {\it post-post-circular} approximation (order $\varepsilon^2$), 
introduced in Ref.~\cite{Damour:2007yf} and then used systematically
in Ref.~\cite{Damour:2007vq,Damour:2008te,Damour:2009kr},
consists of: (i) formally 
solving Eq.~\eqref{eob:4} with respect to the explicit
$p_{\varphi}^2$ appearing on the r.h.s., (ii) replacing $p_{r_*}$ by
its post-circular approximation, Eq.~(\ref{p_adiab}), (iii) using the 
chain rule $d\pi_1(r)/dt = (d\pi_1(r)/dr)(dr/dt)$, and (iv) replacing
$dr/dt$ in terms of $\pi_1$ (to leading order) by using Eq.~\eqref{eob:2}.
The result yields an explicit expression of the type 
$p_\varphi^2 \simeq j_0^2(r)[1 + \varepsilon^2 k_2(r)]$ of which
one finally takes the square root. In principle, this procedure can
be iterated to get initial data at any order in $\varepsilon$.
As it will be shown below, the post-post-circular initial 
data $(j_0\sqrt{1+ \varepsilon^2k_2},\pi_1)$ are sufficient 
to lead to negligible eccentricity when starting the integration 
of the EOB equations of motion at radius $r\equiv R/(GM)=15$.

\subsection{Effective One Body waveforms}
\label{waves}

At this stage we have essentially discussed all the elements that are
needed to compute the EOB dynamics obtained by solving the EOB equation
of motion, Eqs.~\eqref{eob:1}-\eqref{eob:4}.
The dynamics of the system yields a trajectory 
$({\bm q}(t),{\bm p}(t))\equiv (\varphi(t),r(t),p_{\varphi}(t),p_{r_*}(t))$ 
in phase space. The (multipolar) metric waveform during the inspiral
and plunge phase, up to the EOB ``merger time'' $t_m$ 
(that is defined as the maximum of the orbital frequency $\Omega$,) 
is a function of this trajectory, 
i.e. $h_{\lm}^{\rm insplunge}\equiv h_{\lm}^{\rm insplunge}\left({\bm q}(t),{\bm p}(t)\right)$.
Focusing only on the dominant $\ell=m=2$ waveform, the waveform that 
describes the full process of the binary black hole coalescence (i.e.,
inspiral, plunge, merger and ringdown) can be split in two parts:
\begin{itemize}
\item[$\bullet$] The {\it insplunge waveform}:  $h^{\rm insplunge} (t)$,
computed along the EOB dynamics up to merger, which includes 
(i) the resummation of the ``tail'' terms described above and (ii) some 
effective parametrization of Next-to-Quasi-Circular effects.
The $\ell=m=2$ metric waveform explicitly reads
\begin{equation}
\label{eq:h22}
\left( \frac{R c^2}{GM} \right) h_{22}^{\rm insplunge} (t) = \nu
n_{22}^{(0)}c_{2}(\nu)x \,\hat{h}_{22}(\nu;\, x) f_{22}^{\rm NQC} 
Y^{2,-2}\left(\dfrac{\pi}{2},\Phi\right) \, ,
\end{equation}
where the argument $x$ is taken to be (following~\cite{Damour:2006tr}) 
$x=v_{\varphi}^2=(r_{\omega} \Omega)^2$ (where $r_{\omega}$ was
introduced in Eq.~\eqref{eqn13} above).
The resummed version of $f_{22}$ entering in $\hat{h}_{22}(x)$ 
used here is given by the following
{\it Pad\'e-resummed} function $f_{22}^{\rm Pf}\equiv P^3_2[f_{22}^{\rm Taylor}(x;\nu)]$.
In the waveform $h_{22}$ above we have introduced
(following~\cite{Damour:2009kr}) a new ingredient,
a ``Next-to-Quasi-Circular'' (NQC) correction 
factor of the form\footnote{Note that one could also 
similarly improve the subleading higher-multipolar-order 
contributions to ${\mathcal F}_{\varphi}$. In addition, other
(similar) expressions of the NQC factors can be found 
in the literature~\cite{Damour:2007vq,Damour:2008te,Buonanno:2009qa}.}
\begin{equation}
\label{fNQC}
f_{22}^{\rm NQC} (a_1 , a_2) = 1 + a_1 \, \dfrac{p_{r_*}^2}{(r\Omega)^2} 
+ a_2 \, \dfrac{\ddot r}{ r \, \Omega^2} \, ,
\end{equation}
where $a_1$ and $a_2$ are free parameters that have to be fixed.
A crucial facet of the new EOB formalism presented here
consists in trying to be as predictive as possible by
reducing to an absolute minimum the
number of ``flexibility parameters'' entering our theoretical 
framework. One can achieve this aim by ``analytically'' 
determining  the two parameters $a_1, a_2$
entering (via the NQC factor Eq.~\eqref{fNQC})
the (asymptotic) quadrupolar EOB waveform 
$\hat R h^{\rm EOB}_{22}$ (where $\hat R =R/M$)
by imposing: (a) that the modulus $|\hat R h^{\rm EOB}_{22}|$
reaches, at the EOB-determined ``merger time'' $t_m$, a
{\it local maximum}, and (b) that the value of this
maximum EOB modulus is equal to a certain 
(dimensionless) function of $\nu$, $\varphi(\nu)$.
In Ref.~\cite{Damour:2009kr} we  calibrated $\varphi(\nu)$ 
(independently of the EOB formalism) by extracting from the 
best current Numerical Relativity simulations 
the maximum value of the modulus of the 
Numerical Relativity quadrupolar
{\it metric} waveform $|\hat R h^{\rm NR}_{22}|$.
Using the data reported in \cite{Scheel:2008rj} and
\cite{Damour:2008te}, and considering the 
``Zerilli-normalized'' asymptotic metric 
waveform $\Psi_{22} = \hat R h_{22}/\sqrt{24}$, we found
$\varphi(\nu) \simeq 0.3215 \nu ( 1 - 0.131 (1-4\nu))$.
Our requirements (a) and (b) impose,
for any given $A(u)$ potential, {\it two constraints} on the
{\it two parameters} $a_1, a_2$. We can solve these two
constraints (by an iteration procedure) and thereby uniquely
determine the values of $a_1, a_2$ corresponding to any
given $A(u)$ potential. In particular, in the case
considered here where $A(u)\equiv A(u;a_5,a_6,\nu)$ 
this uniquely determines $a_1, a_2$
in function of $a_5,a_6$ and $\nu$.
Note that this is done while also consistently
using  the ``improved'' version 
of $h_{22}$ given by Eq.~\eqref{eq:h22}  
to compute the radiation reaction force via Eq.~\eqref{eq:flux_1}.

\item[$\bullet$] a simplified representation of the transition between plunge 
and ring-down by smoothly {\it matching} (following
Refs.~\cite{Damour:2007xr}), 
on a $(2p+1)$-toothed ``comb'' $(t_m - p\delta , \ldots , t_m - \delta , t_m , t_m + \delta , \ldots
, t_m + p\delta)$ centered around a matching time $t_m$, the inspiral-plus-plunge waveform to a 
ring-down waveform, made of the superposition of 
several\footnote{Refs.~\cite{Damour:2007xr,Damour:2007vq} use $p=2$, {\it
    i.e.} 
a 5-teethed comb, and, correspondingly, 5 positive-frequency Kerr Quasi-Normal
Modes.} 
quasi-normal-mode complex frequencies,
\begin{equation}
\label{eqn17}
\left( \frac{R c^2}{GM} \right) h_{22}^{\rm ringdown} (t) = \sum_N C_N^+ \, e^{-\sigma_N^+ (t-t_m)} \, ,
\end{equation}
with $\sigma_N^+ = \alpha_N + i \, \omega_N$, and where the label $N$ refers
to 
indices $(\ell , \ell' , m , n)$, with $(\ell , m) = (2,2)$ being the 
Schwarzschild-background multipolarity of the considered (metric) waveform 
$h_{\ell m}$, with $n=0,1,2\ldots$ being the `overtone number' of the 
considered Kerr-background Quasi-Normal-Mode, and $\ell'$ the degree of 
its associated spheroidal harmonics $S_{\ell ' m} (a \sigma , \theta)$. 
As discussed in \cite{Buonanno:2000ef} and \cite{Damour:2007xr},
and already mentioned above, the 
physics of the transition between plunge and ring-down (which was first 
understood in the classic work of Davis, Ruffini and 
Tiomno \cite{Davis:1972ud}) suggests to choose as matching time $t_m$, 
in the comparable-mass case, the EOB time when the EOB orbital 
frequency $\Omega (t)$ reaches its {\it maximum} value.
\end{itemize}

Finally, one defines a complete, quasi-analytical EOB waveform 
(covering the full process from inspiral to ring-down) as:
\begin{equation}
\label{eqn18}
h_{22}^{\rm EOB} (t) = \theta (t_m - t) \, h_{22}^{\rm insplunge} (t) + \theta (t-t_m) \, h_{22}^{\rm ringdown} (t) \, ,
\end{equation}
where $\theta (t)$ denotes Heaviside's step function. 
The final result is a waveform that only
depends on the {\it two}  parameters $(a_5,a_6)$
which parametrize some flexibility on the Pad\'e resummation
of the basic radial potential $A(u)$, connected to the
yet uncalculated (4PN, 5PN and) higher PN contributions.

\subsection{Effective One Body dynamics}
\label{sec:eob_evolution}
\begin{figure}[t]
\centering
\includegraphics[height=12.5 cm, width=11.5 cm]{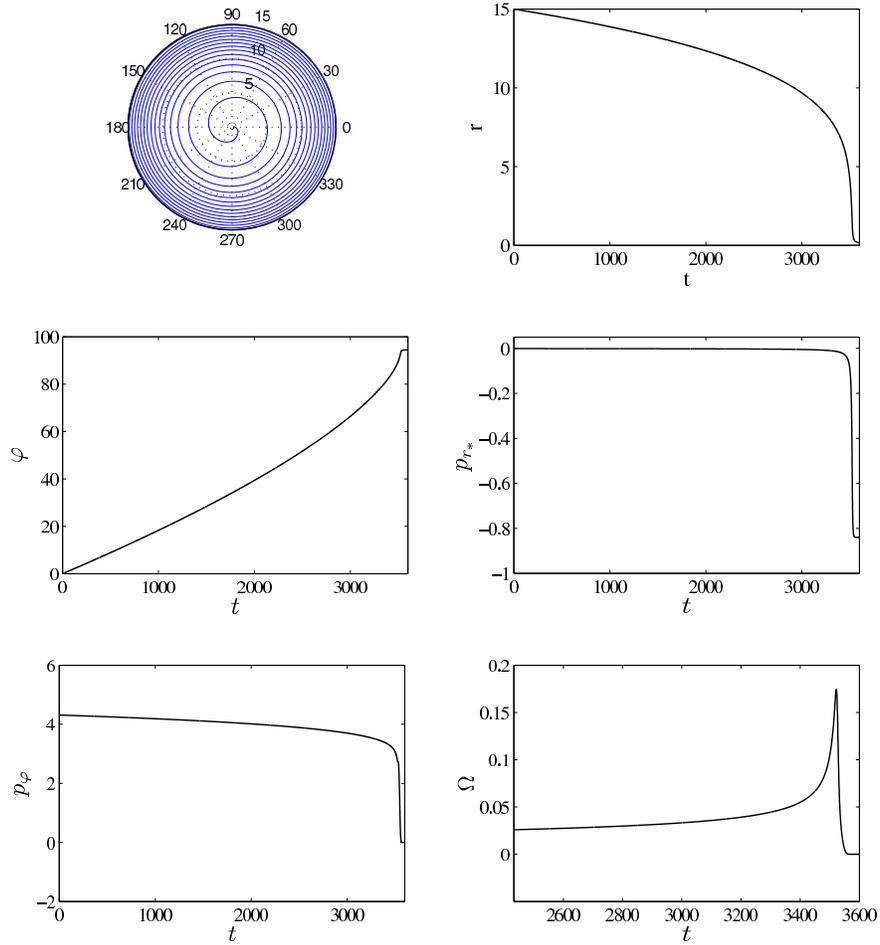}
\caption{EOB dynamics for $a_5=0$ and $a_6=-20$. Clockwise from the top 
left panel,the panels report: the trajectory, the radial separation $r(t)$, 
the radial momentum $p_{r_*}$ (conjugate to $r_*$), the orbital frequency 
$\Omega(t)$, the angular momentum $p_{\varphi}(t)$ and the orbital phase $\varphi(t)$.}
\label{eob_dynamics}       
\end{figure}

We conclude this section by discussing the features of 
the typical EOB dynamics obtained by solving
the EOB equation of motion Eqs.~\eqref{eob:1}-\eqref{eob:4} 
with post-post-circular initial data. 
The resummation of the radiation reaction force uses the multiplicative
decomposition of $h_{\lm}$ given by Eq.~\eqref{eq:hlm} with NQC
correction to the $\ell=m=2$ multipole given by Eq.~\eqref{fNQC}.
We fix the free parameters to the model to be $a_5=0$, $a_6=-20$
(see below why) while $a_1$ and $a_2$ are obtained consistently 
according to the iteration procedure discussed above.
The system is started at $r_0=15$ and $\varphi_0=0$. 
The post-post-circular initial data give 
$p_{\varphi}^0 =4.31509298$ and $p_{r_*}^0=-0.00109847$.
The result of the outcome of the integration
of the EOB equation of motion is displayed in 
Fig.~\ref{eob_dynamics} together with the trajectory 
(top-left panel) and the orbital frequency (bottom-right panel).
On this plot we remark two things. First, the fact that the 
orbital frequency has a maximum at time $t_m=3522$ that 
identifies, in EOB, the merger (and matching) time. 
Second, the fact that $p_{r_{*}}$ tends to a finite value 
after the merger (contrary to $p_r$, that would diverge), 
yielding a more controllable numerical treatment of 
the late part of the EOB dynamics. 

\section{Effective One Body and Numerical Relativity waveforms}
\label{sec:6}

So far we have seen that (at least) two different EOB models
(of dynamics and waveforms) are available. They differ, essentially,
in the way the resummation of the GW energy flux yielding the 
radiation reaction force is performed.
The first EOB model, that we will refer to as the ``old'' one, 
basically uses a Pad\'e-resummation of the energy flux with an
external parameter $v_{\rm pole}$ that must be fixed in some
way. The second EOB model, that we will refer to as the ``improved''
one, uses a more sophisticated resummation procedure of the energy
flux, multipole by multipole, in such a way that the final result
depends explicitly only on the same parameters $(a_5,a_6)$ that
are used to parametrize higher PN contribution to the conservative
part of the dynamics.

In the last three years, the power of the ``old'' EOB model has been
exploited in various comparisons with numerical relativity data,
aiming at constraining in some way the space of the EOB 
flexibility parameters (notably represented by $a_5$ and $v_{\rm pole}$)
 by looking at regions in the parameter space where the agreement between 
the numerical and analytical waveforms is at the level of numerical error.
For example, after a preliminary comparison done in Ref.~\cite{Buonanno:2006ui}, 
Buonanno et al.~\cite{Buonanno:2007pf} compared {\it restricted} 
EOB waveforms\footnote{The terminology ``restricted'' 
refers to a waveform which uses only the leading {\it Newtonian}
approximation, $h_{\lm}^{(N,\epsilon)}$, to the waveform}
to NR waveforms computed by the NASA-Goddard group,
showing that it is possible to tune the value of $a_5$ 
so as to have a good agreement between the two set of data.
In particular, for $a_5=60$ and $v_{\rm pole}$ given according
to the (nowadays outdated) suggestion of Ref.~\cite{Damour:1997ub}, 
in the equal-mass case ($\nu = 1/4$), they found that the dephasing 
between (restricted) EOB and NR waveforms (covering late inspiral, 
merger and ring-down) stayed within $\pm 0.030$ GW cycles 
over 14 GW cycles. In the case of a mass ratio $4:1$ ($\nu = 0.16$), 
the dephasing stayed within $\pm 0.035$ GW cycles over 9 GW cycles. 

Later, the {\it resummed} factorized EOB waveform of Eq.~\eqref{eq:h22} 
above within the ``old'' EOB model has been compared to several set 
of equal-mass and unequal-mass NR waveforms: (i) in the comparison with 
the very accurate inspiralling simulation of the Caltech-Cornell 
group \cite{Boyle:2007ft} the dephasing stayed smaller than $\pm 0.001$ 
GW cycles over 30 GW cycles (and the amplitudes agreed at 
the $\sim 10^{-3}$ level) \cite{Damour:2007yf}; (ii) in 
the comparison~\cite{Damour:2007vq} with a late-inspiral-merger-ringdown NR waveform 
computed by the Albert Einstein Institute group, the dephasing 
stayed smaller than $\pm 0.005$ GW cycles over 12 GW cycles;
(iii) in the (joint) comparison~\cite{Damour:2008te} between EOB and 
very accurate equal-mass inspiralling simulation of the Caltech-Cornell
group~\cite{Boyle:2007ft} and late-inspiral-merger-ringdown
waveform for 1:1, 2:1 and 4:1 mass ratio data computed by the Jena
group it was possible to tune the EOB flexibility parameters
(notably $a_5$ and $v_{\rm pole}$) so that the dephasing stayed
at the level of the numerical error.
The same ``old'' model, with resummed factorized waveform,
and the parameter-dependent (using $v_{\rm pole}$) resummation of 
radiation reaction force, was recently extended by adding 6
more flexibility parameters to the ones already intrduced 
in Refs.~\cite{Damour:2007yf,Damour:2008te}, and was ``calibrated'' on the 
high-accuracy Caltech-Cornell equal-mass data~\cite{Buonanno:2009qa}.
This calibration showed that only 5 flexibility parameters 
($a_5$, $v_{\rm pole}$ and three parameters related to 
non-quasi-circular corrections to the waveform amplitude) 
actually suffice to make the ``old'' EOB and NR waveform 
agree, both in amplitude and phase, at the level of the numerical 
error (this multi-flexed EOB model brings in an improvement with
respect to the one of Refs.~\cite{Damour:2007yf,Damour:2008te}
especially for what concerns the agreement between the waveform
amplitude around the merger).
\begin{figure}[t]
\begin{center}
\includegraphics[height=6cm, width=13 cm]{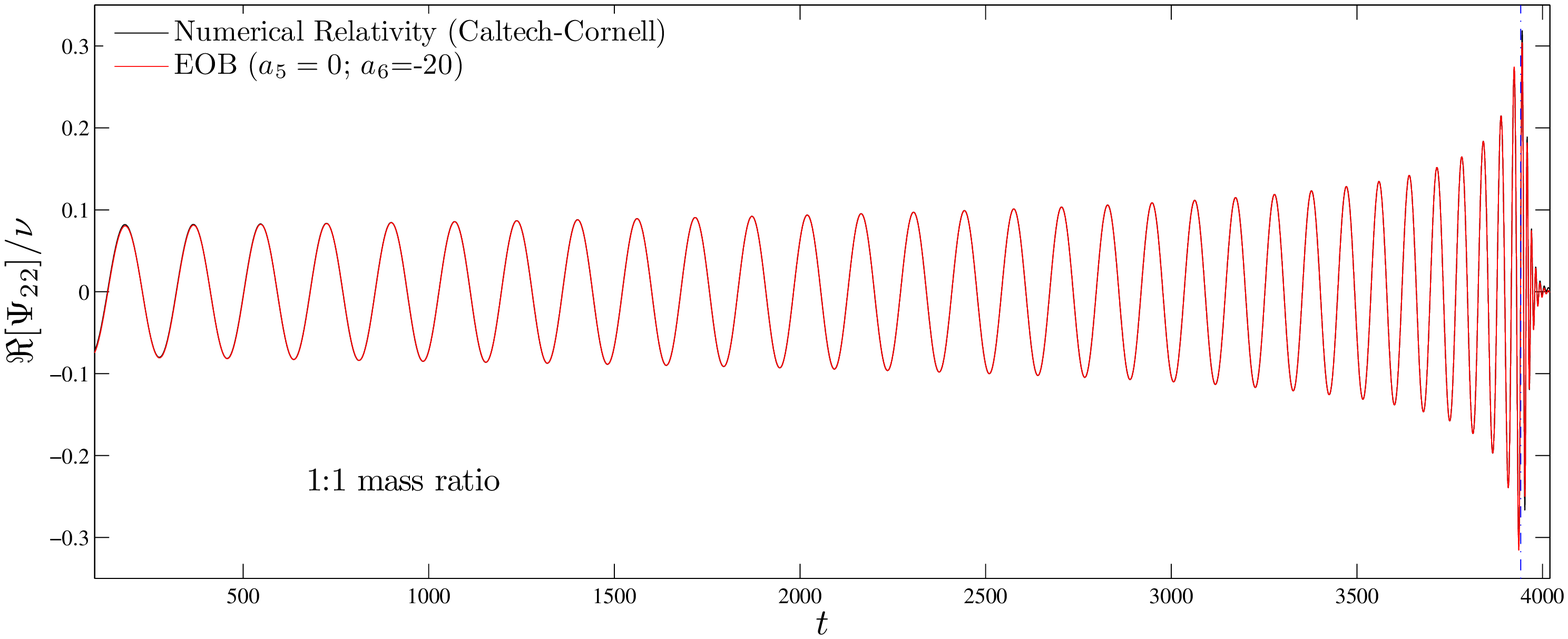}
\caption{This figure illustrates the comparison between 
the ``improved'' EOB waveform 
(quadrupolar ($\ell=m=2$) metric waveform ~\eqref{eq:h22} with 
parameter-free radiation reaction~\eqref{eq:RR_new} and with 
$a_5=0$,$a_6=-20$)
with the most accurate numerical relativity waveform (equal-mass case) nowadays
available. The phase difference between the two is 
$\Delta\phi\leq\pm 0.01$ radians during the entire inspiral
and plunge.  Ref.~\cite{Damour:2009kr} has shown that this
agreement is at the level of the numerical 
error.\label{fig:waveform}}
\end{center}
\end{figure}

Recently, Ref.~\cite{Damour:2009kr} has introduced and fully exploited the
possibilities of the ``improved'' EOB formalism described above, 
taking advantage of: (i) the multiplicative
decomposition of the (resummed) multipolar waveform advocated
in Eq.~\eqref{eq:hlm} above, (ii) the effect of the  NQC 
corrections to the waveform (and energy flux) given by Eq.~\eqref{eq:h22},
and, most importantly, (iii) the parameter-free resummation
of radiation reaction ${\cal F}_{\varphi}$.
\begin{figure}[t]
\begin{center}
\includegraphics[height=8cm ]{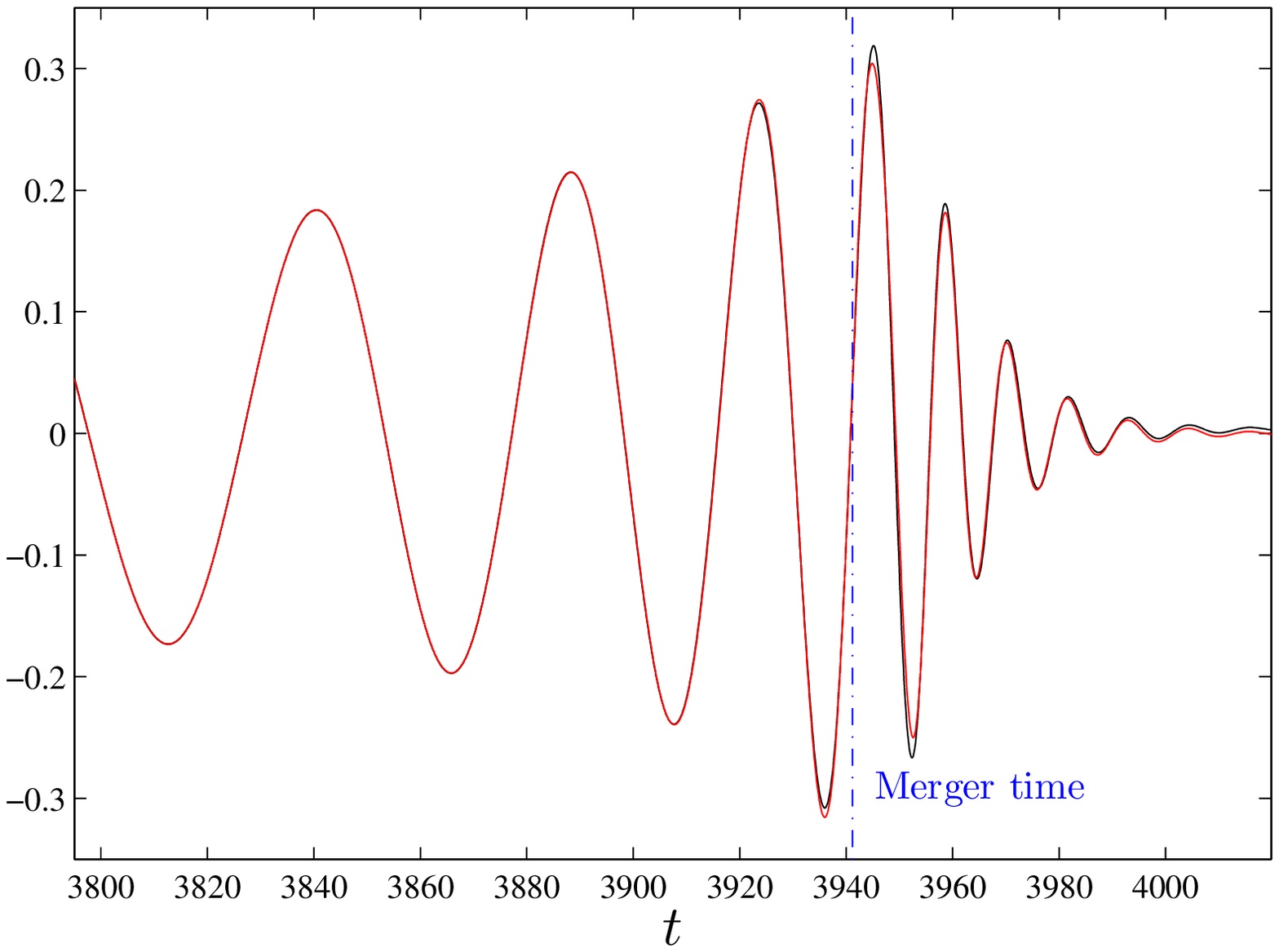}
\caption{\label{fig:ringdown}Close up around merger 
of the waveforms of Fig.~\ref{fig:waveform}. Note the excellent
agreement between {\it both} modulus and phasing also during
the ringdown phase.}
\end{center}
\end{figure}
In Ref.~\cite{Damour:2009kr} the $(a_5,a_6)$-dependent 
predictions made by the ``improved'' formalism were compared to 
the high-accuracy waveform from an equal-mass BBH ($\nu=1/4$) 
computed by the Caltech-Cornell group~\cite{Scheel:2008rj},
(and now made available on the web).
It was found that there is a strong degeneracy between $a_5$ and $a_6$
in the sense that there is an excellent EOB-NR agreement 
for an extended region in the
$(a_5,a_6)$-plane. More precisely, the phase difference between
the EOB (metric) waveform and the Caltech-Cornell one, considered
between GW frequencies $M\omega_{\rm L}=0.047$ and 
$M\omega_{\rm R}=0.31$ (i.e., the last 16 GW cycles before merger),
stays smaller than 0.02 radians within a long and thin 
banana-like region in the $(a_5,a_6)$-plane. This ``good region'' 
approximately extends between the points 
$(a_5,a_6)=(0,-20)$ and $(a_5,a_6)=(-36,+520)$. 
As an example (which actually lies on the boundary of the 
``good region''), we have followed~\cite{Damour:2009kr} 
in considering here the specific values $a_5=0, a_6=-20$ 
(to which correspond, when $\nu=1/4$, $a_1 = -0.036347, a_2=1.2468$). 
We henceforth use $M$ as time unit.

This result relies on the proper comparison between
NR and EOB time series, which is a delicate subject.
In fact, to compare the NR and EOB phase time-series 
$ \phi_{22}^{\rm NR}(t_{\rm NR})$ and $\phi_{22}^{\rm EOB}(t_{\rm EOB})$
one needs to shift, by additive constants, both one of the time variables, and 
one of the phases. In other words, we need to determine $\tau$ and $\alpha$ such that
the ``shifted'' EOB quantities 
\begin{equation}
t'_{\rm EOB}=t_{\rm EOB} + \tau \ , \quad
\phi_{22}^{'\rm EOB} = \phi_{22}^{\rm EOB} + \alpha
\end{equation}
``best fit'' the NR ones. One convenient way to do so is first to ``pinch''
the EOB/NR phase difference at two different instants (corresponding to two
different frequencies). More precisely, one can choose 
two NR times  $t_1^{\rm NR},t_2^{\rm NR}$, which determine  two corresponding
 GW frequencies\footnote{
Alternatively, one can start by giving oneself $\omega_1,\omega_2$ and determine the NR
instants $t_1^{\rm NR},t_2^{\rm NR}$ at which they are reached.}
$\omega_1= \omega_{22}^{\rm NR}(t_1^{\rm NR})$, $\omega_2=\omega_{22}^{\rm NR}(t_2^{\rm NR})$,
and then find the time shift $\tau(\omega_1,\omega_2)$ such that the shifted EOB
phase difference, between $\omega_1$ and $\omega_2$,
$\Delta\phi^{\rm EOB}(\tau) \equiv 
\phi_{22}^{'\rm EOB}(t_2^{'\rm EOB}) - \phi_{22}^{'\rm EOB}(t_1^{'\rm EOB})=
\phi_{22}^{\rm EOB}(t_2^{\rm EOB}+ \tau) - \phi_{22}^{\rm EOB}(t_1^{\rm EOB}+\tau)$
is equal to the corresponding (unshifted) NR phase difference
$\Delta\phi^{\rm NR} \equiv \phi_{22}^{\rm NR}(t_2^{\rm NR}) - \phi_{22}^{\rm NR}(t_1^{\rm NR})$. This yields one equation for one unknown ($\tau$), and (uniquely)
determines a value $\tau(\omega_1,\omega_2)$ of $\tau$. [Note that the 
$\omega_2 \to \omega_1 =\omega_m$ limit of this procedure yields the 
one-frequency matching procedure used in~\cite{Boyle:2007ft}.] After having
so determined $\tau$, one can uniquely define a corresponding best-fit phase shift 
$\alpha(\omega_1,\omega_2)$ by
requiring that, say, 
$\phi_{22}^{'\rm EOB}(t_1^{'\rm EOB}) \equiv \phi_{22}^{\rm EOB}(t_1^{'\rm EOB})
+ \alpha = \phi_{22}^{\rm NR}(t_1^{\rm NR})$.

\begin{figure}[t]
\begin{center}
\includegraphics[height=8cm ]{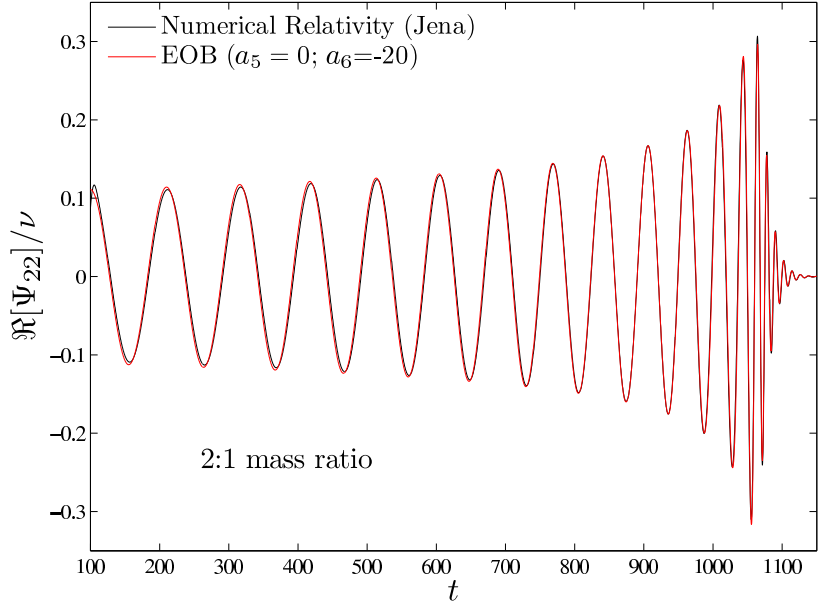}
\caption{\label{fig:fig2_to_1}Comparison between Numerical Relativity 
and EOB metric waveform for the 2:1 mass ratio.}
\end{center}
\end{figure}
Having so related the EOB time and phase variables to the NR ones we can 
straigthforwardly compare all the EOB time series to their NR correspondants.
In particular, we can compute the (shifted) EOB--NR phase difference
\begin{equation}
\label{deltaphi}
\Delta^{\omega_1,\omega_2}\phi_{22}^{\rm EOB NR} (t_{\rm NR}) \equiv \phi_{22}^{'\rm EOB}(t'^{\rm EOB}) - \phi_{22}^{\rm NR}(t^{\rm NR}).
\end{equation}
Figure~\ref{fig:waveform} compares\footnote{The two frequencies used for
this comparison, by means of the ``two-frequency pinching technique''
mentioned above, are $M\omega_1=0.047$ and $M_{\omega_2}=0.31$.} 
(the real part of) our analytical {\it metric} quadrupolar waveform 
$\Psi^{\rm EOB}_{22}/\nu$ to the corresponding 
(Caltech-Cornell) NR {\it metric} waveform 
$\Psi^{\rm NR}_{22}/\nu$. This NR  metric waveform has
been obtained by a double time-integration 
(following the procedure of Ref.~\cite{Damour:2008te})
from the original, publicly available, {\it curvature}
waveform $\psi_4^{22}$. Such a curvature waveform has
been extrapolated {\it both} in resolution and in
extraction radius.
The agreement between the analytical prediction 
and the NR result is striking, even around the merger.
See Fig.~\ref{fig:ringdown} which closes up on the
merger. The vertical line indicates the
location of the EOB-merger time, i.e., the location
of the maximum of the orbital frequency.

\begin{figure}[t]
\begin{center}
\includegraphics[height=8cm]{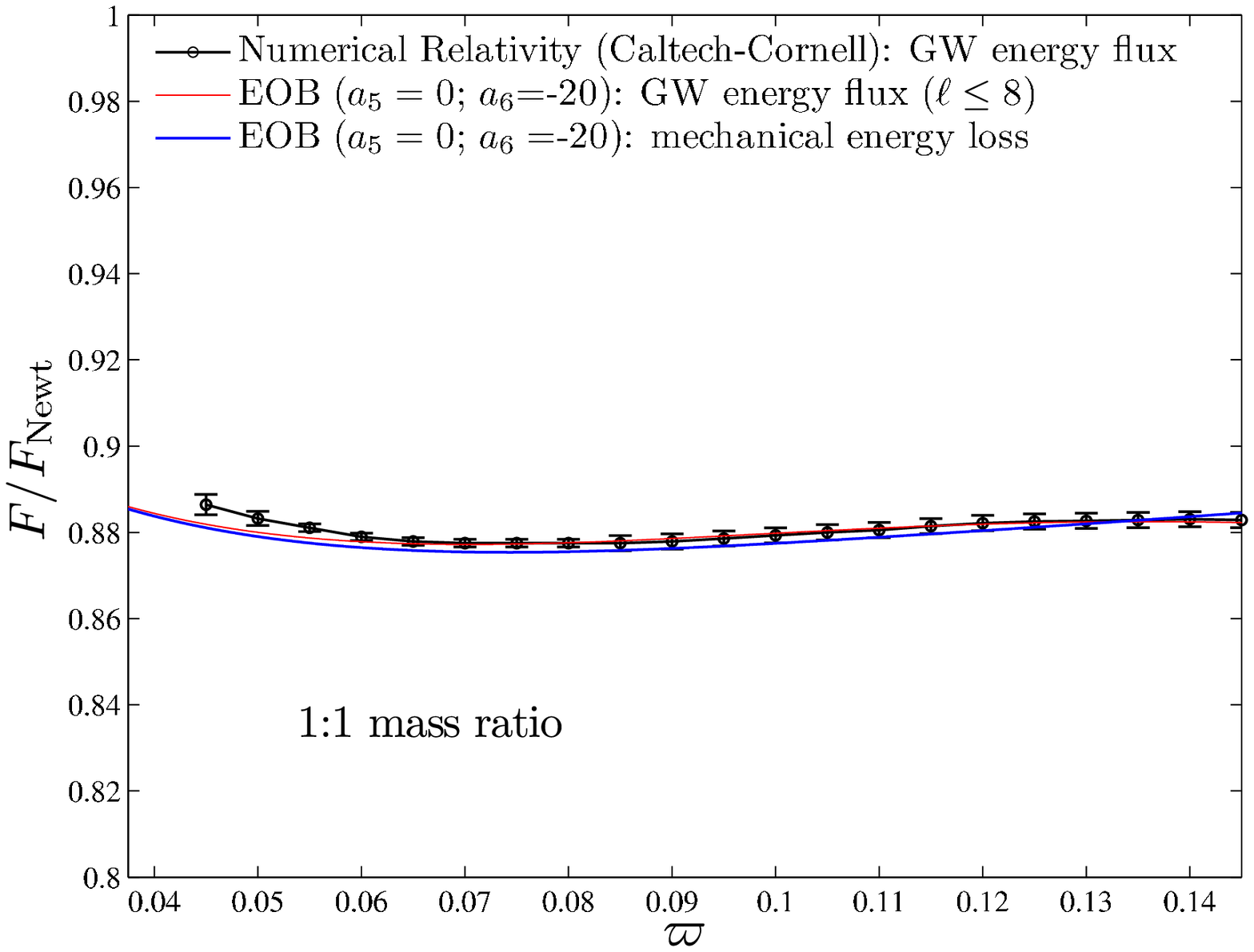}
\caption{ \label{fig:fig4} The triple comparison between 
Numerical Relativity  and EOB GW energy fluxes and 
the EOB mechanical energy loss. }
\end{center}
\end{figure}

The phasing agreement between the waveforms is excellent 
over the full time span of the simulation
(which covers 32 cycles of inspiral and about 6
cycles of ringdown), while the modulus agreement is excellent
over the full span, apart from two cycles after merger
where one can notice a difference. 
More precisely, the phase difference, 
$\Delta \phi= \phi_{\rm metric}^{\rm EOB}-\phi_{\rm metric}^{\rm NR}$, 
remains remarkably small ($\sim \pm 0.02$ radians) during the
entire inspiral and plunge ($\omega_2=0.31$ being quite near
the merger). By comparison, the root-sum of 
the various numerical errors on the phase 
(numerical truncation, outer boundary, 
extrapolation to infinity) is about $0.023$ 
radians during the inspiral~\cite{Scheel:2008rj}. 
At the merger,
and during the ringdown, $\Delta \phi$ takes somewhat 
larger values ($\sim \pm 0.1$ radians), but it oscillates around
zero, so that, on average, it stays very well in phase
with the NR waveform (as is clear on Fig.~\ref{fig:ringdown}). 
By comparison, we note that~\cite{Scheel:2008rj} mentions 
that the phase error linked to the extrapolation to infinity
doubles during ringdown.
We then note that the total ``two-sigma'' NR error level
estimated in~\cite{Scheel:2008rj} rises to
$ 0.05$ radians during ringdown, which is comparable 
to the EOB-NR phase disagreement.
In addition, Ref.~\cite{Damour:2009kr} compared the ``improved''
EOB waveform to accurate numerical relativity data (obtained by
the Jena group~\cite{Damour:2008te}) on the coalescence of
{\it unequal mass-ratio} black-hole binaries. Fig.~\ref{fig:fig2_to_1}
shows the result of the EOB/NR waveform comparison for a 
2:1 mass ratio, corresponding to $\nu=2/9$. When $a_5=0$, $a_6=-20$
one finds $a_1=-0.017017$ and $a_2=1.1906$. Again, the agreement 
is excellent, and within the numerical error bars. 

Finally, Ref.~\cite{Damour:2009kr} explored another aspect of 
the physical soundness of the EOB analytical 
formalism: the {\it triple} comparison
between (i) the NR GW energy flux at infinity
(which was computed in~\cite{Boyle:2008ge}); 
(ii) the corresponding analytically predicted 
GW energy flux at infinity (computed by summing 
$|{\dot h}_{\ell m}|^2$ over $\ell, m$ );
and (iii) (minus) the {\it mechanical} energy loss of the system,
as predicted by the general EOB formalism, i.e. the ``work''
done by the radiation reaction ${\dot E}_{\rm mechanical}
= \Omega {\mathcal F}_{\varphi}$. This comparison is
shown in Fig.~\ref{fig:fig4}, which should be compared 
to Fig. 9 of~\cite{Boyle:2008ge}. 
We kept here the same vertical scale as~\cite{Boyle:2008ge} 
which compared the NR flux to older versions of 
(resummed and non-resummed) analytical fluxes 
and needed such a $\pm 10 \%$ vertical scale
to accomodate all the models they considered. 
[The horizontal axis is the frequency $\varpi$
of the differentiated metric waveform $\dot{h}_{22}$.] 
By contrast, we see again
the striking closeness (at the $ \sim 2 \times 10^{-3}$ level)
between the EOB and NR GW fluxes. As both fluxes
include higher multipoles than the $(2,2)$ one, this
closeness is a further test of the agreement between
the improved EOB formalism and NR results. [We think that
the $\sim 2 \sigma$ difference between the
(coinciding) analytical curves and the NR one on
the left of the Figure is due to uncertainties in the
flux computation of~\cite{Boyle:2008ge}, possibly related 
to the method used there of computing $\dot{h}$.]
Note that the rather close agreement between the analytical
energy flux and the mechanical energy loss during late
inspiral is not required by physics (because of the well-known
``Schott term''~\cite{Schott}), but is rather an 
indication that $\dot h_{\ell m}$ can be well
approximated by $- im \Omega h_{\ell m}$ 

\section{Conclusions}

We have reviewed the basic elements of the Effective One Body (EOB)
formalism. This formalism is still under development. 
The various existing versions of the EOB formalism have all shown 
their capability to reproduce within numerical errors the currently
most accurate numerical relativity simulations of coalescing
binary black holes. These versions differ in the number of free
theoretical parameters.
Recently a new ``improved'' version of the formalism has been defined
which contains essentially only {\it two} free theoretical parameters. 

\noindent Among the successes of the EOB formalism let us mention: 
\begin{enumerate}
\item An analytical understanding of the non-adiabatic late-inspiral 
dynamics and of its ``blurred'' transition to a quasi-circular plunge;
\item The surprising possibility to analytically describe the merger
of two black holes by a seemingly coarse approximation consisting of
matching a continued inspiral to a ringdown signal;
\item The capability, after using suitable resummation methods, to 
reproduce with exquisite accuracy {\it both} the phase and the 
amplitude of the gravitational wave signal emitted during the entire
coalescence process, from early-inspiral,  to late-inspiral, plunge, 
merger and ringdown;
\item The gravitational wave energy flux predicted by the EOB
formalism agrees, within numerical errors, with the most
accurate numerical-relativity energy flux;
\item The ability to correctly estimate (within a 2\% error) 
the final spin and mass of nonspinning coalescing black hole
binaries [this issue has not been discussed in this review,
but see Ref.~\cite{Damour:2007cb}].
\end{enumerate}

We anticipate that the EOB formalism will also
be able to provide an accurate description of more complicated systems
than the nonspinning BBH discussed in this review.
On the one hand,  we think that the recently improved
EOB framework can be extended to the description of
(nearly circularized) {\it spinning} black hole 
systems by suitably incorporating both the PN-expanded
knowledge of spin effects~\cite{Faye:2006gx,Damour:2007nc,Arun:2008kb}
and their possible EOB resummation~\cite{Damour:2001tu,Damour:2008qf}.
On the other hand, the EOB formalism can also be extended
to the description of binary neutron stars or mixed binary
systems made of a black hole and a neutron star~\cite{DN, DNV}.
An important input for this extension is the use of the
relativistic tidal properties of 
neutron stars~\cite{Hinderer:2007mb,Damour:2009vw,Binnington:2009bb}

Finally, we think that the EOB formalism  
has opened the realistic possibility
of constructing (with minimal computational resources) a
very accurate, large bank of gravitational wave templates,
thereby helping in both detecting and analyzing the
signals emitted by inspiralling and coalescing binary
black holes. 
Though we have had in mind in this review essentially 
ground-based detectors, we think that the EOB method 
can also be applied to space-based ones,i.e., to (possibly
eccentric) large mass ratio systems. 

\smallskip

\noindent {\bf Acknowledgments.} 

AN is grateful to Alessandro Spallicci, Bernard Whiting and all 
the organizers of the ``Ecole th\'ematique
du CNRS sur la masse (origine, mouvement, mesure)''.
Among the many colleagues whom we benefitted from,
we would like to thank particularly Emanuele Berti, 
Bernd Br\"ugmann, Alessandra Buonanno, Nils Dorband, 
Mark Hannam, Sascha Husa, Bala Iyer,
Larry Kidder, Eric Poisson, Denis Pollney, 
Luciano Rezzolla, B.S. Sathyaprakash,
Angelo Tartaglia and Loic Villain, for 
fruitful collaborations and discussions.
We are also grateful to Marie-Claude Vergne for help
with Fig.~1.




\newpage 

\begin{thebibliography}{999}
 
\bibitem{Sathyaprakash:2009xs}
  B.~S.~Sathyaprakash and B.~F.~Schutz,
  Physics, Astrophysics and Cosmology with Gravitational Waves,
  Living Rev.\ Rel.\  {\bf 12}, 2 (2009)
  [arXiv:0903.0338 [gr-qc]].


\bibitem{Buonanno:1998gg}
  A.~Buonanno and T.~Damour,
  Effective one-body approach to general relativistic two-body dynamics,
  Phys.\ Rev.\  D {\bf 59}, 084006 (1999)
  [arXiv:gr-qc/9811091].

\bibitem{Buonanno:2000ef}
  A.~Buonanno and T.~Damour,
  Transition from inspiral to plunge in binary black hole coalescences,
  Phys.\ Rev.\  D {\bf 62}, 064015 (2000)
  [arXiv:gr-qc/0001013].

\bibitem{Damour:2001tu}
 T.~Damour,
 Coalescence of two spinning black holes: An effective one-body  approach,
 Phys.\ Rev.\  D {\bf 64}, 124013 (2001)
 [arXiv:gr-qc/0103018].

\bibitem{Damour:2000we}
 T.~Damour, P.~Jaranowski and G.~Sch\"afer,
 On the determination of the last stable orbit for circular general
 relativistic binaries at the third post-Newtonian approximation,
 Phys.\ Rev.\  D {\bf 62}, 084011 (2000)
 [arXiv:gr-qc/0005034].

\bibitem{Buonanno:2005xu}
 A.~Buonanno, Y.~Chen and T.~Damour,
 Transition from inspiral to plunge in precessing binaries of spinning black holes,
 Phys.\ Rev.\  D {\bf 74}, 104005 (2006)
 [arXiv:gr-qc/0508067].

\bibitem{Pretorius:2005gq}
  F.~Pretorius,
  Evolution of Binary Black Hole Spacetimes,
  Phys.\ Rev.\ Lett.\  {\bf 95}, 121101 (2005)
  [arXiv:gr-qc/0507014].

\bibitem{Pretorius:2006tp}
  F.~Pretorius,
  Simulation of binary black hole spacetimes with a harmonic evolution
  scheme,
  Class.\ Quant.\ Grav.\  {\bf 23}, S529 (2006)
  [arXiv:gr-qc/0602115].

\bibitem{Sperhake:2008ga}
  U.~Sperhake, V.~Cardoso, F.~Pretorius, E.~Berti and J.~A.~Gonzalez,
  The high-energy collision of two black holes,
  Phys.\ Rev.\ Lett.\  {\bf 101}, 161101 (2008)
  [arXiv:0806.1738 [gr-qc]].


\bibitem{Campanelli:2005dd}
  M.~Campanelli, C.~O.~Lousto, P.~Marronetti and Y.~Zlochower,
  Accurate Evolutions of Orbiting Black-Hole Binaries Without Excision,
  Phys.\ Rev.\ Lett.\  {\bf 96}, 111101 (2006)
  [arXiv:gr-qc/0511048].

\bibitem{Campanelli:2006uy}
  M.~Campanelli, C.~O.~Lousto and Y.~Zlochower,
  Gravitational radiation from spinning-black-hole binaries: The orbital
  hang up,
  Phys.\ Rev.\  D {\bf 74}, 041501 (2006)
  [arXiv:gr-qc/0604012].

\bibitem{Campanelli:2006gf}
  M.~Campanelli, C.~O.~Lousto and Y.~Zlochower,
  The last orbit of binary black holes,
  Phys.\ Rev.\  D {\bf 73}, 061501 (2006)
  [arXiv:gr-qc/0601091].

\bibitem{Campanelli:2007cga}
  M.~Campanelli, C.~O.~Lousto, Y.~Zlochower and D.~Merritt,
  Maximum gravitational recoil,
  Phys.\ Rev.\ Lett.\  {\bf 98}, 231102 (2007)
  [arXiv:gr-qc/0702133].

\bibitem{Baker:2005vv}
  J.~G.~Baker, J.~Centrella, D.~I.~Choi, M.~Koppitz and J.~van Meter,
  Gravitational wave extraction from an inspiraling configuration of  merging
  black holes,
  Phys.\ Rev.\ Lett.\  {\bf 96}, 111102 (2006)
  [arXiv:gr-qc/0511103].

\bibitem{Baker:2006yw}
  J.~G.~Baker, J.~Centrella, D.~I.~Choi, M.~Koppitz and J.~van Meter,
  Binary black hole merger dynamics and waveforms,
  Phys.\ Rev.\  D {\bf 73}, 104002 (2006)
  [arXiv:gr-qc/0602026].

\bibitem{Baker:2006vn}
  J.~G.~Baker, J.~Centrella, D.~I.~Choi, M.~Koppitz, J.~R.~van Meter and M.~C.~Miller,
  Getting a kick out of numerical relativity,
  Astrophys.\ J.\  {\bf 653}, L93 (2006)
  [arXiv:astro-ph/0603204].

\bibitem{Baker:2007fb}
  J.~G.~Baker, M.~Campanelli, F.~Pretorius and Y.~Zlochower,
  Comparisons of binary black hole merger waveforms,
  Class.\ Quant.\ Grav.\  {\bf 24}, S25 (2007)
  [arXiv:gr-qc/0701016].

\bibitem{Gonzalez:2006md}
  J.~A.~Gonzalez, U.~Sperhake, B.~Bruegmann, M.~Hannam and S.~Husa,
  Total recoil: the maximum kick from nonspinning black-hole binary
  inspiral,
  Phys.\ Rev.\ Lett.\  {\bf 98}, 091101 (2007)
  [arXiv:gr-qc/0610154].

\bibitem{Gonzalez:2007hi}
  J.~A.~Gonzalez, M.~D.~Hannam, U.~Sperhake, B.~Bruegmann and S.~Husa,
  Supermassive kicks for spinning black holes,
  Phys.\ Rev.\ Lett.\  {\bf 98}, 231101 (2007)
  [arXiv:gr-qc/0702052].

\bibitem{Husa:2007rh}
  S.~Husa, M.~Hannam, J.~A.~Gonzalez, U.~Sperhake and B.~Bruegmann,
  Reducing eccentricity in black-hole binary evolutions with initial
  parameters from post-Newtonian inspiral,
  Phys.\ Rev.\  D {\bf 77}, 044037 (2008)
  [arXiv:0706.0904 [gr-qc]].


\bibitem{Koppitz:2007ev}
  M.~Koppitz, D.~Pollney, C.~Reisswig, L.~Rezzolla, J.~Thornburg, P.~Diener and E.~Schnetter,
  Getting a kick from equal-mass binary black hole mergers,
  Phys.\ Rev.\ Lett.\  {\bf 99}, 041102 (2007)
  [arXiv:gr-qc/0701163].

\bibitem{Pollney:2007ss}
  D.~Pollney {\it et al.},
  Recoil velocities from equal-mass binary black-hole mergers: a systematic
  investigation of spin-orbit aligned configurations,
  Phys.\ Rev.\  D {\bf 76}, 124002 (2007)
  [arXiv:0707.2559 [gr-qc]].

\bibitem{Rezzolla:2007xa}
  L.~Rezzolla, E.~N.~Dorband, C.~Reisswig, P.~Diener, D.~Pollney, E.~Schnetter and B.~Szilagyi,
  Spin Diagrams for Equal-Mass Black-Hole Binaries with Aligned Spins,
  Astrophysics {\bf J679}, 1422 (2008)
  [arXiv:0708.3999 [gr-qc]].

\bibitem{Rezzolla:2007rd}
  L.~Rezzolla, P.~Diener, E.~N.~Dorband, D.~Pollney, C.~Reisswig, E.~Schnetter and J.~Seiler,
  The final spin from the coalescence of aligned-spin black-hole binaries,
  Astrophys.\ J.\  {\bf 674}, L29 (2008)
  [arXiv:0710.3345 [gr-qc]].

\bibitem{Rezzolla:2007rz}
  L.~Rezzolla, E.~Barausse, E.~N.~Dorband, D.~Pollney, C.~Reisswig, J.~Seiler and S.~Husa,
  On the final spin from the coalescence of two black holes,
  Phys.\ Rev.\  D {\bf 78}, 044002 (2008)
  [arXiv:0712.3541 [gr-qc]].


\bibitem{Boyle:2006ne}
  M.~Boyle, L.~Lindblom, H.~Pfeiffer, M.~Scheel and L.~E.~Kidder,
  Testing the Accuracy and Stability of Spectral Methods in Numerical
  Relativity,
  Phys.\ Rev.\  D {\bf 75}, 024006 (2007)
  [arXiv:gr-qc/0609047].

\bibitem{Boyle:2007ft}
  M.~Boyle {\it et al.},
  High-accuracy comparison of numerical relativity simulations with
  post-Newtonian expansions,
  Phys.\ Rev.\  D {\bf 76}, 124038 (2007)
  [arXiv:0710.0158 [gr-qc]].

\bibitem{Boyle:2008ge}
  M.~Boyle, A.~Buonanno, L.~E.~Kidder, A.~H.~Mroue, Y.~Pan, H.~P.~Pfeiffer and M.~A.~Scheel,
  High-accuracy numerical simulation of black-hole binaries: Computation of
  the gravitational-wave energy flux and comparisons with post-Newtonian
  approximants,
  Phys.\ Rev.\  D {\bf 78}, 104020 (2008)
  [arXiv:0804.4184 [gr-qc]].

\bibitem{Scheel:2008rj}
  M.~A.~Scheel, M.~Boyle, T.~Chu, L.~E.~Kidder, K.~D.~Matthews and H.~P.~Pfeiffer,
  High-accuracy waveforms for binary black hole inspiral, merger, and
  ringdown,
  Phys.\ Rev.\  D {\bf 79}, 024003 (2009)
  [arXiv:0810.1767 [gr-qc]].

\bibitem{Pretorius:2007nq}
  F.~Pretorius,
  Binary Black Hole Coalescence . 
  The final version of this Lecture Note will
  appear in the book: {\it Relativistic Objects in Compact Binaries: From
  Birth to Coalescense}, M.~Colpi et al. Eds.,Springer Verlag, Canopus
  Publishing Limited,  arXiv:0710.1338 [gr-qc]

\bibitem{Buonanno:2006ui}
  A.~Buonanno, G.~B.~Cook and F.~Pretorius,
  Inspiral, merger and ring-down of equal-mass black-hole binaries,
  Phys.\ Rev.\  D {\bf 75}, 124018 (2007)
  [arXiv:gr-qc/0610122].

\bibitem{Pan:2007nw}
  Y.~Pan {\it et al.},
  A data-analysis driven comparison of analytic and numerical coalescing
  binary waveforms: Nonspinning case,
  Phys.\ Rev.\  D {\bf 77}, 024014 (2008)
  [arXiv:0704.1964 [gr-qc]].

\bibitem{Buonanno:2007pf}
  A.~Buonanno, Y.~Pan, J.~G.~Baker, J.~Centrella, B.~J.~Kelly, S.~T.~McWilliams and J.~R.~van Meter,
  Toward faithful templates for non-spinning binary black holes using the
  effective-one-body approach,
  Phys.\ Rev.\  D {\bf 76}, 104049 (2007)
  [arXiv:0706.3732 [gr-qc]].

\bibitem{Damour:2007cb}
  T.~Damour and A.~Nagar,
  Final spin of a coalescing black-hole binary: An effective-one-body
  approach,
  Phys.\ Rev.\  D {\bf 76}, 044003 (2007)
  [arXiv:0704.3550 [gr-qc]].

\bibitem{Nagar:2006xv}
  A.~Nagar, T.~Damour and A.~Tartaglia,
  Binary black hole merger in the extreme mass ratio limit,
  Class.\ Quant.\ Grav.\  {\bf 24}, S109 (2007)
  [arXiv:gr-qc/0612096].


\bibitem{Damour:2007xr}
  T.~Damour and A.~Nagar,
  Faithful Effective-One-Body waveforms of small-mass-ratio coalescing
  black-hole binaries,
  Phys.\ Rev.\  D {\bf 76}, 064028 (2007)
  [arXiv:0705.2519 [gr-qc]].

\bibitem{Damour:2007yf}
  T.~Damour and A.~Nagar,
  Comparing Effective-One-Body gravitational waveforms to accurate numerical
  data,
  Phys.\ Rev.\  D {\bf 77}, 024043 (2008)
  [arXiv:0711.2628 [gr-qc]].

\bibitem{Damour:2007vq}
 T.~Damour, A.~Nagar, E.~N.~Dorband, D.~Pollney and L.~Rezzolla,
 Faithful Effective-One-Body waveforms of equal-mass coalescing black-hole
 binaries,
 Phys.\ Rev.\  D {\bf 77}, 084017 (2008)
 [arXiv:0712.3003 [gr-qc]].

\bibitem{Damour:2008te}
 T.~Damour, A.~Nagar, M.~Hannam, S.~Husa and B.~Bruegmann,
 Accurate Effective-One-Body waveforms of inspiralling and coalescing
 black-hole binaries,
 Phys.\ Rev.\  D {\bf 78}, 044039 (2008)
 [arXiv:0803.3162 [gr-qc]].

\bibitem{Damour:2009kr}
 T.~Damour and A.~Nagar,
 An improved analytical description of inspiralling and coalescing
 black-hole binaries,
 arXiv:0902.0136 [gr-qc].

\bibitem{Buonanno:2009qa}
  A.~Buonanno, Y.~Pan, H.~P.~Pfeiffer, M.~A.~Scheel, L.~T.~Buchman and L.~E.~Kidder,
  Effective-one-body waveforms calibrated to numerical relativity
  simulations: coalescence of non-spinning, equal-mass black holes,
  arXiv:0902.0790 [gr-qc].

\bibitem{Baker:2006ha}
  J.~G.~Baker, J.~R.~van Meter, S.~T.~McWilliams, J.~Centrella and B.~J.~Kelly,
  Consistency of post-Newtonian waveforms with numerical relativity,
  Phys.\ Rev.\ Lett.\  {\bf 99}, 181101 (2007)
  [arXiv:gr-qc/0612024

\bibitem{Hannam:2007ik}
  M.~Hannam, S.~Husa, U.~Sperhake, B.~Bruegmann and J.~A.~Gonzalez,
  Where post-Newtonian and numerical-relativity waveforms meet,
  Phys.\ Rev.\  D {\bf 77}, 044020 (2008)
  [arXiv:0706.1305 [gr-qc]].


\bibitem{Ajith:2007qp}
  P.~Ajith {\it et al.},
  Phenomenological template family for black-hole coalescence waveforms,
  Class.\ Quant.\ Grav.\  {\bf 24}, S689 (2007)
  [arXiv:0704.3764 [gr-qc]].

\bibitem{Ajith:2007kx}
  P.~Ajith {\it et al.},
  A template bank for gravitational waveforms from coalescing binary black
  holes: I. non-spinning binaries,
  Phys.\ Rev.\  D {\bf 77}, 104017 (2008)
  [arXiv:0710.2335 [gr-qc]].

\bibitem{Gopakumar:2007vh}
  A.~Gopakumar, M.~Hannam, S.~Husa and B.~Bruegmann,
  Comparison between numerical relativity and a new class of post-Newtonian
  gravitational-wave phase evolutions: the non-spinning equal-mass case,
  Phys.\ Rev.\  D {\bf 78}, 064026 (2008)
  [arXiv:0712.3737 [gr-qc]].

\bibitem{Hannam:2007wf}
  M.~Hannam, S.~Husa, B.~Bruegmann and A.~Gopakumar,
  Comparison between numerical-relativity and post-Newtonian waveforms from
  spinning binaries: the orbital hang-up case,
  Phys.\ Rev.\  D {\bf 78}, 104007 (2008)
  [arXiv:0712.3787 [gr-qc]].


\bibitem{Brezin:1970zr}
 E.~Brezin, C.~Itzykson and J.~Zinn-Justin,
 Relativistic balmer formula including recoil effects,
 Phys.\ Rev.\  D {\bf 1}, 2349 (1970).

\bibitem{Damour:1997ub}
 T.~Damour, B.~R.~Iyer and B.~S.~Sathyaprakash,
 Improved filters for gravitational waves from inspiralling compact
 binaries,
 Phys.\ Rev.\  D {\bf 57}, 885 (1998)
 [arXiv:gr-qc/9708034].

\bibitem{Damour:2008gu}
 T.~Damour, B.~R.~Iyer and A.~Nagar,
 Improved resummation of post-Newtonian multipolar waveforms from
 circularized compact binaries,
 arXiv:0811.2069 [gr-qc].

\bibitem{Davis:1972ud}
 M.~Davis, R.~Ruffini and J.~Tiomno,
 Pulses of gravitational radiation of a particle falling radially into a
 Schwarzschild black hole,
 Phys.\ Rev.\  D {\bf 5}, 2932 (1972).

\bibitem{Price:1994pm}
 R.~H.~Price and J.~Pullin,
 Colliding black holes: The Close limit,
 Phys.\ Rev.\ Lett.\  {\bf 72}, 3297 (1994)
 [arXiv:gr-qc/9402039].

\bibitem{Damour:1981bh}
  T.~Damour and N.~Deruelle,
  Radiation Reaction And Angular Momentum Loss In Small Angle Gravitational
  Scattering,
  Phys.\ Lett.\  A {\bf 87}, 81 (1981).

\bibitem{Damour:1982wm}
  T.~Damour, Gravitational Radiation And The Motion Of Compact Bodies,,
  in {\it Gravitational Radiation}, N.~Deruelle and T.~Piran Eds, 1983, North
  Holland, Amsterdam, p 59-144.

\bibitem{Schafer:1986rd}
  G.~Sch\"afer,
  The gravitational quadrupole radiation reaction force and the 
  canonical formalism of ADM,
  Annals Phys.\  {\bf 161}, 81 (1985).

\bibitem{kopejkin1985}
  S.~M.~Kopejkin: Astron. Zh. {\bf 62}, 889 (1985).


\bibitem{gr-qc/9712075} 
P.~Jaranowski and G.~Sch\"afer,
3rd post-Newtonian higher order Hamilton dynamics for two-body point-mass systems,
Phys.\ Rev.\  D {\bf 57}, 7274 (1998)
[Erratum-ibid.\  D {\bf 63}, 029902 (2001)]
[arXiv:gr-qc/9712075].

\bibitem{gr-qc/0007051} 
 L.~Blanchet and G.~Faye,
 General relativistic dynamics of compact binaries at the third
 post-Newtonian order,
  Phys.\ Rev.\  D {\bf 63}, 062005 (2001)
  [arXiv:gr-qc/0007051].

\bibitem{gr-qc/0105038} 
T.~Damour, P.~Jaranowski and G.~Sch\"afer,
Dimensional regularization of the gravitational interaction of point
masses,
Phys.\ Lett.\  B {\bf 513}, 147 (2001)
[arXiv:gr-qc/0105038].

\bibitem{gr-qc/0311052} 
 L.~Blanchet, T.~Damour and G.~Esposito-Farese,
 Dimensional regularization of the third post-Newtonian dynamics of point
 particles in harmonic coordinates,''
 Phys.\ Rev.\  D {\bf 69}, 124007 (2004)
[arXiv:gr-qc/0311052].

\bibitem{IF03} 
Y.~Itoh and T.~Futamase,
New derivation of a third post-Newtonian equation of motion for
relativistic compact binaries without ambiguity,
Phys.\ Rev.\  D {\bf 68}, 121501 (2003)
[arXiv:gr-qc/0310028].

\bibitem{gr-qc/0201001} 
M.~E.~Pati and C.~M.~Will,
Post-Newtonian gravitational radiation and equations of motion via  direct
integration of the relaxed Einstein equations. II: Two-body  equations of
motion to second post-Newtonian order, and  radiation-reaction to 3.5
post-Newton,
Phys.\ Rev.\  D {\bf 65}, 104008 (2002)
[arXiv:gr-qc/0201001].

\bibitem{KonigsdorfferFayeSchafer03} 
C.~Konigsdorffer, G.~Faye and G.~Sch\"afer,
The binary black-hole dynamics at the third-and-a-half post-Newtonian order
in the ADM-formalism,''
Phys.\ Rev.\  D {\bf 68}, 044004 (2003)
[arXiv:gr-qc/0305048].

\bibitem{gr-qc/0412018} 
S.~Nissanke and L.~Blanchet,
Gravitational radiation reaction in the equations of motion of compact
binaries to 3.5 post-Newtonian order,''
Class.\ Quant.\ Grav.\  {\bf 22}, 1007 (2005)
[arXiv:gr-qc/0412018].

\bibitem{Blanchet:1985sp}
  L.~Blanchet and T.~Damour,
  Radiative gravitational fields in general relativity I. 
  General structure of the field outside the source,
  Phil.\ Trans.\ Roy.\ Soc.\ Lond.\  A {\bf 320}, 379 (1986).

\bibitem{Blanchet:1989ki}
L.~Blanchet and T.~Damour,
Postnewtonian generation of gravitational waves,
Annales Poincare Phys.\ Theor.\  {\bf 50} 377, (1989).

\bibitem{Damour:1990gj}
  T.~Damour and B.~R.~Iyer,
  Multipole analysis for electromagnetism and linearized gravity with
  irreducible cartesian tensors,
  Phys.\ Rev.\  D {\bf 43}, 3259 (1991).

\bibitem{Damour:1990ji}
  T.~Damour and B.~R.~Iyer,
  PostNewtonian generation of gravitational waves. 2. The Spin moments,
  Annales Poincare Phys.\ Theor.\  {\bf 54}, 115 (1991).

\bibitem{Blanchet:1992br}
  L.~Blanchet and T.~Damour,
  Hereditary Effects In Gravitational Radiation,
  Phys.\ Rev.\  D {\bf 46}, 4304 (1992).

\bibitem{gr-qc/9501030} 
L.~Blanchet,
Second Postnewtonian Generation Of Gravitational Radiation,
Phys.\ Rev.\  D {\bf 51}, 2559 (1995)
[arXiv:gr-qc/9501030].

\bibitem{gr-qc/9710038}
 L.~Blanchet,
 Gravitational-wave tails of tails,
 Class.\ Quant.\ Grav.\  {\bf 15}, 113 (1998)
 [Erratum-ibid.\  {\bf 22}, 3381 (2005)]
 [arXiv:gr-qc/9710038].

\bibitem{gr-qc/9608012}  
C.~M.~Will and A.~G.~Wiseman,
Gravitational radiation from compact binary systems: gravitational
waveforms and energy loss to second post-Newtonian order,
Phys.\ Rev.\  D {\bf 54}, 4813 (1996)
[arXiv:gr-qc/9608012].

\bibitem{gr-qc/9910057} 
C.~M.~Will,
Generation of post-Newtonian gravitational radiation via direct
integration of the relaxed Einstein equations,
  Prog.\ Theor.\ Phys.\ Suppl.\  {\bf 136}, 158 (1999)
  [arXiv:gr-qc/9910057].

\bibitem{gr-qc/0007087} 
M.~E.~Pati and C.~M.~Will,
Post-Newtonian gravitational radiation and equations of motion via  direct
integration of the relaxed Einstein equations. I: Foundations,
  Phys.\ Rev.\  D {\bf 62}, 124015 (2000)
  [arXiv:gr-qc/0007087].

\bibitem{gr-qc/9501027}  
L.~Blanchet, T.~Damour, B.~R.~Iyer, C.~M.~Will and A.~G.~Wiseman,
Gravitational Radiation Damping Of Compact Binary Systems To Second
Postnewtonian Order,
  Phys.\ Rev.\ Lett.\  {\bf 74}, 3515 (1995)
  [arXiv:gr-qc/9501027].

\bibitem{gr-qc/9501029} 
L.~Blanchet, T.~Damour and B.~R.~Iyer,
Gravitational Waves From Inspiralling Compact Binaries: Energy Loss And
Wave Form To Second Postnewtonian Order,
  Phys.\ Rev.\  D {\bf 51}, 5360 (1995)
  [Erratum-ibid.\  D {\bf 54}, 1860 (1996)]
  [arXiv:gr-qc/9501029].


\bibitem{gr-qc/0105098} 
L.~Blanchet, B.~R.~Iyer and B.~Joguet,
Gravitational waves from inspiralling compact binaries: Energy flux to
third post-Newtonian order,
  Phys.\ Rev.\  D {\bf 65}, 064005 (2002)
  [Erratum-ibid.\  D {\bf 71}, 129903 (2005)]
  [arXiv:gr-qc/0105098].


\bibitem{gr-qc/0409094} 
L.~Blanchet and B.~R.~Iyer,
Hadamard regularization of the third post-Newtonian gravitational wave
generation of two point masses,
  Phys.\ Rev.\  D {\bf 71}, 024004 (2005)
  [arXiv:gr-qc/0409094].


\bibitem{W93} 
A.~G.~Wiseman,
Coalescing Binary Systems Of Compact Objects To 
(Post)Newtonian5/2 Order.4v: The Gravitational 
Wave Tail,
Phys.\ Rev.\  D {\bf 48}, 4757 (1993).

\bibitem{BS93} 
L.~Blanchet and G.~Sch\"a fer,
Gravitational wave tails and binary star systems,
  Class.\ Quant.\ Grav.\  {\bf 10}, 2699 (1993).


\bibitem{gr-qc/0406012}
L.~Blanchet, T.~Damour, G.~Esposito-Farese and B.~R.~Iyer,
  Gravitational radiation from inspiralling compact binaries completed at
  the third post-Newtonian order,
  Phys.\ Rev.\ Lett.\  {\bf 93}, 091101 (2004)
  [arXiv:gr-qc/0406012].

\bibitem{gr-qc/0503044}
 L.~Blanchet, T.~Damour, G.~Esposito-Farese and B.~R.~Iyer,
 Dimensional regularization of the third post-Newtonian gravitational  wave
 generation from two point masses,
 Phys.\ Rev.\  D {\bf 71}, 124004 (2005)
 [arXiv:gr-qc/0503044].

\bibitem{Blanchet:2002av}
  L.~Blanchet,
  Gravitational radiation from post-Newtonian sources and inspiralling
  compact binaries,
  Living Rev.\ Rel.\  {\bf 5}, 3 (2002)
  [arXiv:gr-qc/0202016].

\bibitem{Damour:2000zb}
  T.~Damour, B.~R.~Iyer and B.~S.~Sathyaprakash,
  A comparison of search templates for gravitational waves from binary inspiral,
  Phys.\ Rev.\  D {\bf 63}, 044023 (2001)
  [Erratum-ibid.\  D {\bf 72}, 029902 (2005)]
  [arXiv:gr-qc/0010009].

\bibitem{gr-qc/0105099}
 L.~Blanchet, G.~Faye, B.~R.~Iyer and B.~Joguet,
 Gravitational-wave inspiral of compact binary systems to 7/2 post-Newtonian order,
 Phys.\ Rev.\  D {\bf 65}, 061501 (2002)
 [Erratum-ibid.\  D {\bf 71}, 129902 (2005)]
 [arXiv:gr-qc/0105099].

\bibitem{Baker:2006kr}
  J.~G.~Baker, S.~T.~McWilliams, J.~R.~van Meter, J.~Centrella, D.~I.~Choi, B.~J.~Kelly and M.~Koppitz,
  Binary black hole late inspiral: Simulations for gravitational wave observations,
  Phys.\ Rev.\  D {\bf 75}, 124024 (2007)
  [arXiv:gr-qc/0612117].

\bibitem{Damour:1988mr}
 T.~Damour and G.~Sch\"afer,
 Higher order relativistic periastron advances and binary pulsars,
 Nuovo Cim.\  B {\bf 101}, 127 (1988).

\bibitem{gr-qc/0211041}
 T.~Damour, B.~R.~Iyer, P.~Jaranowski and B.~S.~Sathyaprakash,
 Gravitational waves from black hole binary inspiral and merger: The span of
 third post-Newtonian effective-one-body templates,
 Phys.\ Rev.\  D {\bf 67}, 064028 (2003)
 [arXiv:gr-qc/0211041].

\bibitem{Damour:2002qh}
  T.~Damour, E.~Gourgoulhon and P.~Grandclement,
  Circular orbits of corotating binary black holes: Comparison between
  analytical and numerical results,
  Phys.\ Rev.\  D {\bf 66}, 024007 (2002)
  [arXiv:gr-qc/0204011].

\bibitem{Damour:2006tr}
 T.~Damour and A.~Gopakumar,
 Gravitational recoil during binary black hole coalescence using the
 effective one body approach,
 Phys.\ Rev.\  D {\bf 73}, 124006 (2006)
 [arXiv:gr-qc/0602117].

\bibitem{gr-qc/9405062} 
H.~Tagoshi and M.~Sasaki,
PostNewtonian expansion of gravitational waves from a particle in circular
orbit around a Schwarzschild black hole,
  Prog.\ Theor.\ Phys.\  {\bf 92}, 745 (1994)
  [arXiv:gr-qc/9405062].

\bibitem{Cutler:1993vq}
 C.~Cutler, E.~Poisson, G.~J.~Sussman and L.~S.~Finn,
 Gravitational radiation from a particle in circular orbit around a black
 hole. 2: Numerical results for the nonrotating case,
 Phys.\ Rev.\  D {\bf 47}, 1511 (1993).

\bibitem{Yunes:2008tw}
 N.~Yunes and E.~Berti,
 Accuracy of the Post-Newtonian Approximation: Optimal Asymptotic Expansion
 for Quasi-Circular, Extreme-Mass Ratio Inspirals,
 Phys.\ Rev.\  D {\bf 77}, 124006 (2008)
 [arXiv:0803.1853 [gr-qc]].

\bibitem{Poisson:1995vs}
 E.~Poisson,
 Gravitational radiation from a particle in circular orbit around a black
 hole. 6. Accuracy of the postNewtonian expansion,
 Phys.\ Rev.\  D {\bf 52}, 5719 (1995)
 [Addendum-ibid.\  D {\bf 55}, 7980 (1997)]
 [arXiv:gr-qc/9505030].

\bibitem{Tagoshi:1994sm}
 H.~Tagoshi and M.~Sasaki,
 PostNewtonian expansion of gravitational waves from a particle in circular
 orbit around a Schwarzschild black hole,
 Prog.\ Theor.\ Phys.\  {\bf 92}, 745 (1994)
 [arXiv:gr-qc/9405062].

\bibitem{Tanaka:1997dj}
 T.~Tanaka, H.~Tagoshi and M.~Sasaki,
 Gravitational waves by a particle in circular orbits around a
 Schwarzschild black hole: 5.5 post-Newtonian formula,
 Prog.\ Theor.\ Phys.\  {\bf 96}, 1087 (1996)
 [arXiv:gr-qc/9701050].

\bibitem{Damour:2002vi}
  T.~Damour, B.~R.~Iyer, P.~Jaranowski and B.~S.~Sathyaprakash,
  Gravitational waves from black hole binary inspiral and merger: The span of
  third post-Newtonian effective-one-body templates,
  Phys.\ Rev.\  D {\bf 67}, 064028 (2003)
  [arXiv:gr-qc/0211041].

\bibitem{gr-qc/9701050} 
T.~Tanaka, H.~Tagoshi and M.~Sasaki, 
Prog. Theor. Phys. {\bf 96} 1087 (1996), 
[arXiv:gr-qc/9701050v1].

\bibitem{Kidder:2007rt}
  L.~E.~Kidder,
  Using Full Information When Computing Modes of Post-Newtonian Waveforms
  From Inspiralling Compact Binaries in Circular Orbit,
  Phys.\ Rev.\  D {\bf 77}, 044016 (2008)
  [arXiv:0710.0614 [gr-qc]].

\bibitem{Berti:2007fi}
  E.~Berti, V.~Cardoso, J.~A.~Gonzalez, U.~Sperhake, M.~Hannam, S.~Husa and B.~Bruegmann,
  Inspiral, merger and ringdown of unequal mass black hole binaries: A
  multipolar analysis,
  Phys.\ Rev.\  D {\bf 76}, 064034 (2007)
  [arXiv:gr-qc/0703053].

\bibitem{Blanchet:2008je}
  L.~Blanchet, G.~Faye, B.~R.~Iyer and S.~Sinha,
  The third post-Newtonian gravitational wave polarisations and associated
  spherical harmonic modes for inspiralling compact binaries in quasi-circular
  orbits,
  arXiv:0802.1249 [gr-qc].

\bibitem{Blanchet:1997ji}
  L.~Blanchet,
  Quadrupole-quadrupole gravitational waves,
  Class.\ Quant.\ Grav.\  {\bf 15}, 89 (1998)
  [arXiv:gr-qc/9710037].

\bibitem{Blanchet:1997jj}
  L.~Blanchet,
  Gravitational-wave tails of tails,
  Class.\ Quant.\ Grav.\  {\bf 15}, 113 (1998)
  [Erratum-ibid.\  {\bf 22}, 3381 (2005)]
  [arXiv:gr-qc/9710038].

\bibitem{Schott}
G.~A.~Schott, Phil. Mag. {\bf 29}, 49, (1915).

\bibitem{Faye:2006gx}
  G.~Faye, L.~Blanchet and A.~Buonanno,
  Higher-order spin effects in the dynamics of compact binaries. I:
  Equations of motion,
  Phys.\ Rev.\  D {\bf 74}, 104033 (2006)
  [arXiv:gr-qc/0605139].

\bibitem{Damour:2007nc}
  T.~Damour, P.~Jaranowski and G.~Sch\"afer,
  Hamiltonian of two spinning compact bodies with next-to-leading order
  gravitational spin-orbit coupling,
  Phys.\ Rev.\  D {\bf 77}, 064032 (2008)
  [arXiv:0711.1048 [gr-qc]].

\bibitem{Damour:2008qf}
  T.~Damour, P.~Jaranowski and G.~Sch\"afer,
  Effective one body approach to the dynamics of two spinning black holes
  with next-to-leading order spin-orbit coupling,
  Phys.\ Rev.\  D {\bf 78}, 024009 (2008)
  [arXiv:0803.0915 [gr-qc]].

\bibitem{Arun:2008kb}
  K.~G.~Arun, A.~Buonanno, G.~Faye and E.~Ochsner,
  Higher-order spin effects in the amplitude and phase of gravitational
  waveforms emitted by inspiraling compact binaries: Ready-to-use gravitational
  waveforms,
  arXiv:0810.5336 [gr-qc].

\bibitem{DN}
T.~Damour and A.~Nagar, in preparation.

\bibitem{DNV}
T.~Damour, A.~Nagar and L.~Villain, in preparation.

\bibitem{Hinderer:2007mb}
T.~Hinderer,
Tidal Love numbers of neutron stars,
Astrophys.\ J.\  {\bf 677}, 1216 (2008)
[arXiv:0711.2420 [astro-ph]].

\bibitem{Damour:2009vw}
T.~Damour and A.~Nagar,
Relativistic tidal properties of neutron stars,
arXiv:0906.0096 [gr-qc].


\bibitem{Binnington:2009bb}
T.~Binnington and E.~Poisson,
Relativistic theory of tidal Love numbers,
arXiv:0906.1366 [gr-qc].

\end{thebibliography}
\end{document}